\documentclass{emulateapj}

\newcommand{\HII}{H {\footnotesize II}} 
 
\newcommand{\HeI}{[He {\footnotesize I}]} 
\newcommand{\NII}{[N {\footnotesize II}]} 
\newcommand{\NIII}{[N {\footnotesize III}]} 
\newcommand{\OII}{[O {\footnotesize II}]} 
\newcommand{\OIII}{[O {\footnotesize III}]} 
\newcommand{\SIII}{[S {\footnotesize III}]} 

\newcommand{\ngc}{NGC~4449}

\newcommand{\flamunits}{erg~s$^{-1}$~cm$^{-2}$~\AA$^{-1}$}
\newcommand{\fluxunits}{erg~s$^{-1}$~cm$^{-2}$}
\newcommand{\hst}{{\it HST}}
\newcommand{\ha}{H$\alpha$}
\newcommand{\hbeta}{H$\beta$}
\newcommand{\av}{A$_{\rm V}$}

\shorttitle{NEBULAR EMISSION IN YOUNG MASSIVE STAR CLUSTERS}
\shortauthors{Reines et al.}

\begin{document}

\title{THE IMPORTANCE OF NEBULAR CONTINUUM AND LINE EMISSION \\
IN OBSERVATIONS OF YOUNG MASSIVE STAR CLUSTERS} 

\author{Amy E. Reines, David L. Nidever, David G. Whelan and Kelsey E. Johnson\altaffilmark{*}}
\affil{Department of Astronomy, University of Virginia, P.O. Box 400325,
    Charlottesville, VA 22904-4325}
\email{areines@virginia.edu}

\altaffiltext{*}{Adjunct at National Radio Astronomy Observatory, 520 Edgemont Road, Charlottesville, VA 22903, USA}

\begin{abstract}

In this spectroscopic study of infant massive star clusters, we find that continuum
emission from ionized gas rivals the stellar luminosity at optical wavelengths.  In addition, we
find that nebular line emission is significant in many commonly used broad-band {\it Hubble Space
Telescope (HST)} filters including the F814W $I$-band, the F555W $V$-band and the F435W $B$-band.
Two young massive clusters (YMCs) in the nearby starburst galaxy \ngc\ were targeted for
follow-up spectroscopic observations after  \citet{Reines08a} discovered an F814W $I$-band excess
in their photometric study of radio-detected clusters in the galaxy.  The spectra were obtained
with the Dual Imaging Spectrograph (DIS) on the 3.5 m Apache Point Observatory (APO)
telescope\footnote{The 3.5 m Apache Point Observatory telescope is owned and operated
by the Astrophysical Research Consortium.} and have a spectral range of $\sim$3800-9800~\AA.
We supplement these data with \hst\ and Sloan Digital Sky Survey (SDSS) photometry
of the clusters.  By comparing our data to the Starburst99 and GALEV evolutionary synthesis models, we
find that nebular continuum emission competes with the stellar light in our observations
and that the relative contribution from the nebular continuum is largest in the $U$- and $I$-bands, where the
Balmer (3646 \AA) and Paschen jumps (8207 \AA) are located.  The spectra also exhibit strong line emission
including the \SIII\ $\lambda \lambda 9069,9532$ lines in the \hst\ F814W $I$-band.  We find that the
{\it combination} of nebular continuum and line emission can account for the F814W $I$-band excess previously
found by \citet{Reines08a}.  In an effort to provide a benchmark for estimating the impact of 
ionized gas emission on photometric observations of young massive stellar populations, we compute the
relative contributions of the stellar continuum, nebular continuum, and emission lines to the
total observed flux of a 3 Myr-old cluster through various \hst\ filter/instrument combinations,
including filters in the Wide Field Camera 3 (WFC3).  We urge caution when comparing observations of YMCs to
evolutionary synthesis models since nebular continuum and line emission can have a large impact on magnitudes
and colors of young ($\lesssim 5$ Myr) clusters, significantly affecting inferred properties such as ages,
masses and extinctions.

\end{abstract}


\section{INTRODUCTION}

Massive star clusters are an important mode of star formation, having
an impact on a wide range of galaxy properties.  However, the earliest
stages of these clusters are notoriously challenging to study
since the youngest clusters are still enshrouded in remnants of their
gaseous and dusty birth cocoons.  Properly accounting for the effects
of gas and dust on observations of young massive clusters (YMCs)
is nontrivial, rendering it difficult to understand their formation and
earliest phases of evolution.  
Some of these difficulties, such as interstellar extinction, are commonly
dealt with in studies of young star clusters.  While it is often not possible
to completely disentangle the effects of extinction, the general effects are
well-known and accounted for \citep[e.g.][]{Calzetti94,Whitmore99,Johnson99,Reines08b}.  

On the other hand, the treatment of ionized gas emission can be quite complicated
and the effects of nebular emission on broad-band observations of YMCs are not
always obvious.  Emission lines detected via spectroscopy or narrow-band imaging,
however, have proved useful for both identifying young ($\lesssim 10$ Myr) clusters
(e.g. \ha) and serving as valuable diagnostics on their physical conditions
\citep[e.g.][]{Eissner69,Castaneda92,Kewley02}.
If these lines are strong enough, as is typically the case for
massive star forming regions, the composite emission can even affect the integrated
broad-band photometry \citep[e.g.][]{Johnson99,Anders03}.  However, nebular emission also takes
on a more subtle form via continuum emission (e.g. free-free, free-bound).  In
fact, discontinuities or ``jumps'' in the nebular continuum (e.g. the Balmer
and Paschen jumps) can be strong enough to serve as diagnostics in their own right
\citep[see][and references therein]{Guseva06}.
This nebular continuum emission can also severely affect the
broad-band photometry of very young clusters \citep[e.g.][]{Leitherer95,Molla09}.

\begin{figure*}
\begin{center}
\includegraphics[scale=0.8]{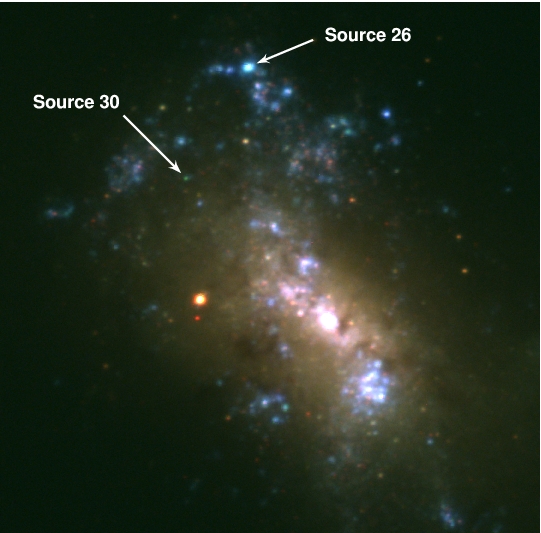}
\caption{An SDSS 3-color image of \ngc\ (RGB=$ugi$) using a logarithmic scaling.  The
two YMCs for which we have spectra are labeled \citep[Sources 26 and 30 according to
the notation of][]{Reines08a}.  The field of view is approximately $3\arcmin \times 3\arcmin$.
\label{image_n4449}}
\end{center}
\end{figure*}

In this paper we investigate the impact of nebular continuum and
line emission on observations of very young massive
star clusters.  We carry out a detailed multiwavelength study of two YMCs that
are still partially embedded in their birth material.  The clusters were targeted
for spectroscopic follow-up after \citet{Reines08a} found an \hst\ F814W $I$-band excess
in their photometric study of radio-detected clusters in the nearby dwarf starburst
galaxy \ngc.  \citet{Reines08a} considered several possible explanations for this
$I$-band excess, including red supergiants, emission lines, thermal emission from hot dust,
continuum emission from a sub-population of deeply embedded stars, and Extended Red
Emission \citep[ERE;][]{Witt04}.  With the observations presented here, we have
determined that the {\it combination} of nebular continuum and line emission
can account for the $I$-band excess found by \citet{Reines08a}.  Consequently,
the chief goal of this study is to gain a better empirical understanding of
the effects of nebular emission (both continuum and line) associated with extremely young
massive star clusters.

This paper is organized as follows:  The data are described in \S\ref{sec_data}.  In
\S\ref{sec_lines} we estimate the physical properties of the clusters from their
\ha\ emission.  We model the spectral energy distributions (SEDs) in \S\ref{sec_model}
and investigate the impact
of nebular emission (line and continuum) on broad-band photometry in \S\ref{sec_broad-band}.
A summary of our main conclusions is given in \S\ref{sec_summary}.

\section{DATA}\label{sec_data}

A three-color SDSS image of \ngc\ is shown in Figure \ref{image_n4449} with our target
YMCs labeled.  Source 26 \citep[according to the notation of][]{Reines08a} was
selected for its brightness (to get a high signal-to-noise spectrum) and Source 30
was targeted for its large $I$-band excess -- the largest of all of the clusters
studied in \citet{Reines08a}.  We adopt a distance of 3.9 Mpc to \ngc\ as in
\citet{Reines08a}, consistent with the results of \citet{Annibali08}.  At this
distance, $1\arcsec \sim 19$~pc. 

\begin{figure*}
\begin{center}
\includegraphics[scale=0.85]{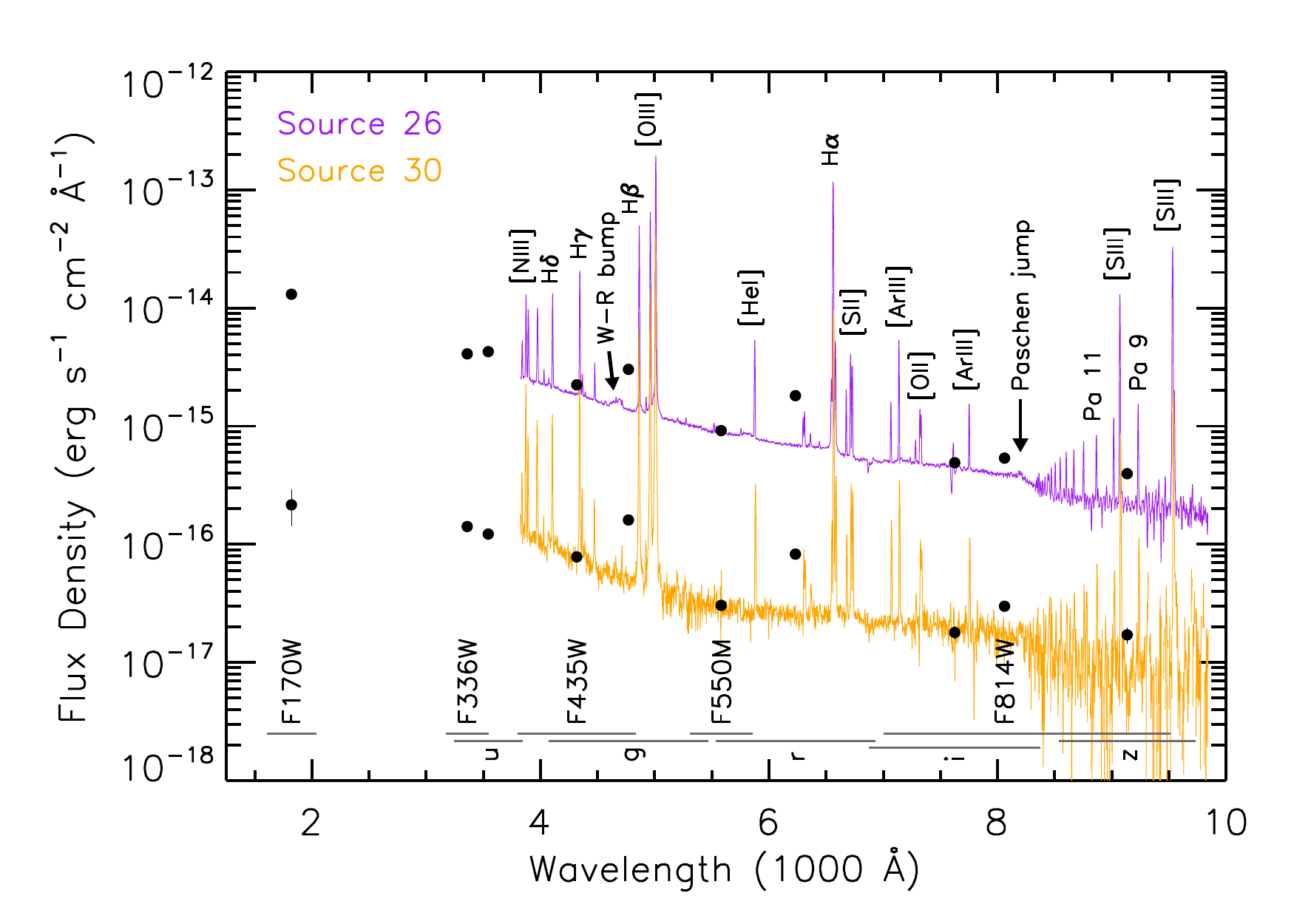}
\caption{APO/DIS flux calibrated spectra of our target YMCs.  \hst\ and SDSS broad-band photometry
are plotted as dots and the approximate widths of the filters are shown as gray bars.
The spectra exhibit many strong emission lines as well as nebular continuum emission
as is evident by the Paschen jump at 8207 \AA.  The ``W-R bump'' at $\sim 4650$ \AA\ signifying the
presence of Wolf-Rayet stars is also clearly visible in the spectrum of Source 26.}\label{plot_spectra}
\end{center}
\end{figure*}

\subsection{Spectroscopy}\label{sec_spec}

Spectra of the two YMCs were obtained using the 3.5~m APO telescope.  Source 26
(RA = 12$^{\rm h}$ 28$^{\rm m}$ 13.86$^{\rm s}$, DEC = +44$\degr$ 07\arcmin\ 10.4\arcsec)
was observed on the night of 2008 April 13 under clear conditions and Source 30
(RA = 12$^{\rm h}$ 28$^{\rm m}$ 16.02$^{\rm s}$, DEC = +44$\degr$ 06\arcmin\ 29.3\arcsec) was
observed on 2009 May 24 through intermittent clouds.  The observations were made using
the red and blue channels of the Dual Imaging Spectrograph (DIS) in the low-resolution
mode with the B400 and R300 gratings.  The linear dispersions for the blue and red
channels are 1.83 and 2.31~\AA\ pix$^{-1}$, respectively, and the central wavelengths
are $\sim$4400 and $\sim$7500~\AA.  We used the $1\farcs5 \times 360\arcsec$ slit
oriented along the parallactic angle and centered on the YMCs.  The resulting spectra
have a wavelength range of $\sim$3800-9800~\AA\ and a resolution of $\sim$7~\AA.
The total exposure time on Source 26 was 30 minutes (2 $\times$ 15 minutes), resulting
in signal-to-noise ratios of $\sim$60 and $\sim$45 in the blue and red continua,
respectively.  Source 30 was observed for a total of 2.5 hours (5 $\times$ 30
minutes) and the signal-to-noise ratio of both the blue and red continua is $\sim$5.
Multiple exposures of each cluster were taken to allow for the removal
of cosmic rays.  The white dwarf primary spectrophotometric standard star GD 153
\citep{Bohlin95} was observed for flux calibration and a He-Ne-Ar arc lamp spectrum was
taken for wavelength calibration.

The data were reduced using a combination of IRAF and custom IDL routines.
The two-dimensional images were processed using standard procedures in IRAF including
bias subtraction and flat-fielding \citep{Massey97}.  IRAF was also used to extract the
one-dimensional spectra, subtract the background, and apply the wavelength calibration
\citep{Massey92}.  The spectra were extracted from an aperture window of 11 pixels
(4\farcs4) in the spatial dimension.  The remaining data reduction and analysis was performed in IDL.
Flux calibration was obtained by comparing the observed spectrum of GD 153 to the model
spectrum obtained from the Space Telescope Science Institute's Calibration Data Base
System.\footnote{http://www.stsci.edu/hst/observatory/cdbs/calspec.html}  Atmospheric extinction
was corrected for using the standard extinction curve for APO.  Finally, the blue and red
spectral regions were combined and put on a common wavelength scale of 2~\AA\ pix$^{-1}$.  
The blue and red spectra were averaged in an overlap region approximately 50 pixels wide
and centered at $\lambda \sim 5370$\AA.

After a comparison with the photometry (see below),
the flux calibrated spectra were multiplied by scale factors of 1.94 (Source 26) and 1.84
(Source 30) to account for loss of light in the slit.  We scaled the spectra to match the
\hst\ F550M (narrow $V$-band) data since this filter has a negligible contribution from emission
lines.  Figure \ref{plot_spectra} shows the final reduced spectra of Sources 26 and 30.
The spectra exhibit many strong emission lines and nebular continuum
emission as is evident by the Paschen jump at 8207 \AA.  The Wolf-Rayet blended emission feature
at $\sim 4650$ \AA\ known as the ``W-R bump'' \citep[e.g.][]{Conti91,Schaerer98} is also clearly visible
in the spectrum of Source 26.

\subsection{Photometry}\label{sec_phot}

% HST Observations
\begin{deluxetable*}{cccccc}
\tabletypesize{\footnotesize}
\tablecolumns{6} 
\tablewidth{0pt} 
\tablecaption{Archival {\it HST} Observations of NGC~4449\label{tab_hstobs}} 
\tablehead{ 
\colhead{Filter}  &  \colhead{Instrument} & \colhead{Description} &
\colhead{Exp. Time (s)} & \colhead{Proposal ID} & \colhead{PI}}
\startdata
F170W & WFPC2   & UV                             & 400 $\times$ 2  & 6716 & T. Stecher \\
F336W & WFPC2   & WFPC2 U                        & 260 $\times$ 2  & 6716 & T. Stecher \\
F435W & ACS/WFC & Johnson B                      & 3660  & 10585 & A. Aloisi         \\
F550M & ACS/WFC & Narrow V                       & 1200  & 10522 & D. Calzetti          \\
F658N & ACS/WFC & H$\alpha$ + [NII]$\lambda6584$ & 1539  & 10585 & A. Aloisi          \\
F660N & ACS/WFC & [NII]$\lambda6584$             & 1860  & 10522 & D. Calzetti         \\
F814W & ACS/WFC & Broad I                        & 2060  & 10585 & A. Aloisi        
\enddata
\end{deluxetable*} 

\hst\ observations of \ngc\ taken with the Advanced Camera for Surveys (ACS) and the
Wide Field and Planetary Camera 2 (WFPC2) were obtained from the archive.
These data are described in detail in \citet{Reines08a} and summarized in Table \ref{tab_hstobs}.
Here we use WFPC2 F170W and F336W images, as well as ACS
Wide Field Channel (WFC) images through the F435W,
F550M, F658N, F660N and F814W filters.  The broad-band filters (F170W, F336W,
F435W, F550M and F814W) span ultraviolet through near-IR wavelengths.  The
narrow-band filters, F658N and F660N, capture \ha + \NII\ $\lambda6584$ and
\NII\ $\lambda6584$ emission, respectively.
We have also obtained $u g r i z$ imaging of \ngc\ from the SDSS Data Release 6 \citep{York00,
Adelman08}.  A three-color (RGB=$ugi$) image of the galaxy is shown in Figure \ref{image_n4449}
with our target YMCs labeled.

\begin{figure}
\begin{center}
\includegraphics[scale=0.47]{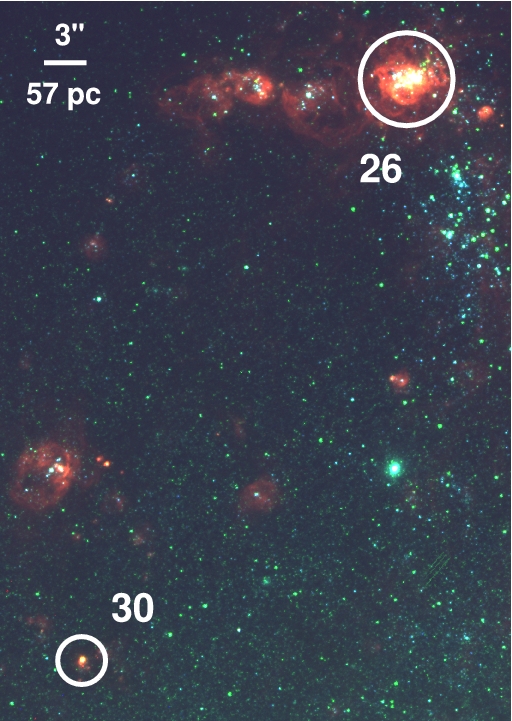}
\caption{An \hst\ 3-color image of \ngc\ in the vicinity of Sources 26 and 30.
Red, green and blue correspond to the F658N (\ha), F814W (broad $I$),
and F550M (narrow $V$) filters.  The circles indicate the photometric apertures
used on the \hst\ and SDSS imaging (see \S\ref{sec_phot}).
\label{hstim}}
\end{center}
\end{figure}

Photometry of the YMCs was performed using SURPHOT \citep{Reines08a}, a custom IDL
program allowing for consistent apertures and background annuli across multiple
wavebands.  Circular apertures of radii 3\farcs3 and 1\farcs7 were used for Sources
26 and 30, respectively.  Background levels were determined in annuli with inner
and outer radii equal to $1.75 \times$ and $2.5\times$ the aperture radii.
Aperture sizes were chosen to include essentially all of the
light from the clusters in all bands, and were mostly influenced by the
(seeing-limited) resolution of the ground-based SDSS images and the intrinsic
cluster sizes.  Figure \ref{hstim} shows a three-color \hst\ image of the
clusters with the photometric apertures overlaid.
Flux densities for the YMCs are given in Table \ref{tab_phot} and plotted in
Figure \ref{plot_spectra} along with the spectra.

% HST Photometry
\begin{deluxetable}{cccc}
\tabletypesize{\footnotesize}
\tablecolumns{4} 
\tablewidth{0pt} 
\tablecaption{Flux Densities of the YMCs\label{tab_phot}} 
\tablehead{ 
\colhead{Filter}  &  \colhead{$\lambda$ (\AA)} & \colhead{Source 26} & \colhead{Source 30}}
\startdata
\cutinhead{\hst}
F170W & 1821 & $1.3(0.1) \times 10^{-14}$ & $2.2(0.7) \times 10^{-16}$ \\
F336W & 3359 & $4.1(0.4) \times 10^{-15}$ & $1.4(0.1) \times 10^{-16}$ \\
F435W & 4317 & $2.2(0.2) \times 10^{-15}$ & $7.8(0.8) \times 10^{-17}$ \\
F550M & 5581 & $9.2(0.9) \times 10^{-16}$ & $3.0(0.3) \times 10^{-17}$ \\
F658N & 6584 & $1.6(0.2) \times 10^{-14}$ & $1.1(0.1) \times 10^{-15}$ \\
F660N & 6599 & $3.4(0.3) \times 10^{-15}$ & $2.6(0.3) \times 10^{-16}$ \\
F814W & 8061 & $5.4(0.5) \times 10^{-16}$ & $3.0(0.3) \times 10^{-17}$ \\
\cutinhead{SDSS}
$u$ & 3543 & $4.3(0.4) \times 10^{-15}$ & $1.2(0.1) \times 10^{-16}$ \\
$g$ & 4770 & $3.0(0.3) \times 10^{-15}$ & $1.6(0.2) \times 10^{-16}$ \\
$r$ & 6231 & $1.8(0.2) \times 10^{-15}$ & $8.3(0.9) \times 10^{-17}$ \\
$i$ & 7625 & $4.9(0.5) \times 10^{-16}$ & $1.8(0.2) \times 10^{-17}$ \\
$z$ & 9134 & $4.0(0.4) \times 10^{-16}$ & $1.7(0.3) \times 10^{-17}$
\enddata
\tablecomments{Units of flux density are
\flamunits.  Uncertainties are shown in parentheses.}
\end{deluxetable} 

\subsection{A Comparison of the Spectroscopic and Photometric Data}\label{sec_comp}

\subsubsection{Broad-band Flux Densities}\label{sec_fluxden}

\begin{figure}
\begin{center}
\includegraphics[scale=0.5]{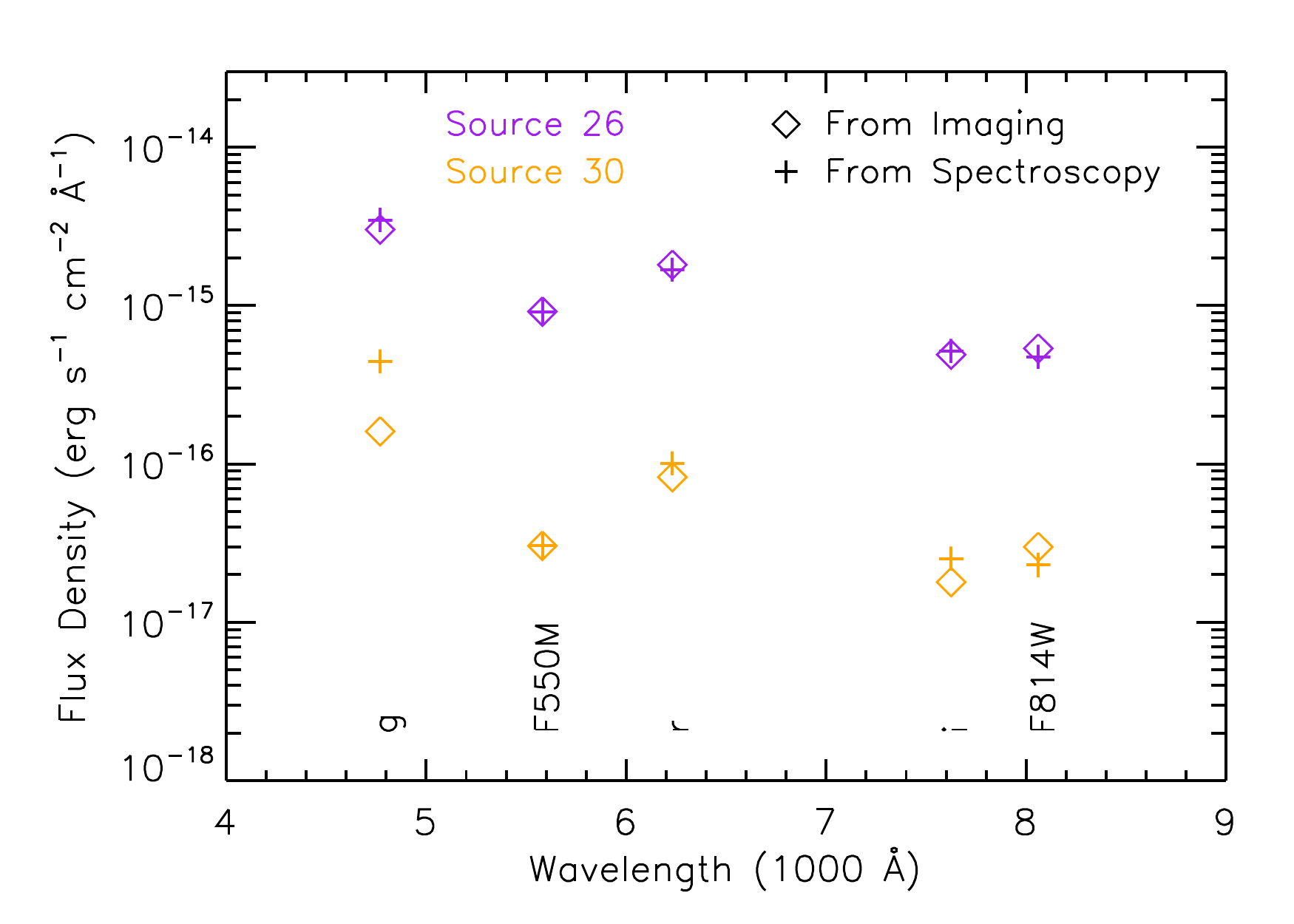}
\caption{Broad-band flux densities obtained from the imaging data (diamonds) and
from convolving the observed spectra with the total system throughput curves for the
filters shown (plus signs).  The sizes of the plot symbols reflect the errors.  The 
spectroscopic data has been scaled to match the F550M photometry to account for loss of
light in the slit.  The discrepancy between the $g$ flux densities for
Source 30 are most likely due to an incorrect absolute flux level in the blue part of the
spectrum (see \S\ref{sec_fluxden}).  }\label{plot_phots}
\end{center}
\end{figure}

In order to compare our spectroscopic observations with the broad-band imaging data, we
simulate ``DIS photometry'' by convolving the YMC spectra with the total system throughput
curves for the $g$, F550M, $r$, $i$, and F814W filters.\footnote{The observed spectra do not
cover the entire wavelength ranges of the F435W and $z$ filters.}
Figure \ref{plot_phots} shows the broad-band flux densities obtained from the
imaging data (listed in Table \ref{tab_phot}), as well as the flux densities
obtained from the spectroscopy.  

Figure \ref{plot_phots} illustrates that the flux densities obtained from convolving the
spectra with the filter throughput curves match the the photometry quite well, especially
for Source 26 which has a high signal-to-noise spectrum.  However, the SDSS $g$ flux densities
for Source 30 are clearly discrepant, with the spectroscopic value significantly higher than
the photometric value.  We believe this is due to an incorrect absolute flux level in the blue
part of the spectrum for two reasons.  First, we expect the F435W photometric data point to lie above
the continuum in Figure \ref{plot_spectra} since this filter contains strong emission lines
(e.g. H$\gamma$ and H$\delta$).  This effect can be seen for Source 26 in Figure \ref{plot_spectra}
(i.e. the F435W data point lies above the continuum in the spectrum),
and the emission lines are even stronger relative to the continuum for Source 30.
The second reason we think the blue end of the spectrum of Source 30 has an incorrect
flux level is that no single model SED can reproduce the shape of the continuum in the spectrum.
There is no combination of age and extinction that can simultaneously produce such a
steep blue spectrum and flattened red spectrum.  Also, we show later in \S\ref{sec_model} that the
best-fitting SED to the broad-band {\it photometric} data has a much flatter slope than the blue part
of the spectrum of Source 30.  

It is not absolutely clear why the spectrum of Source 30 has
an incorrect absolute flux level in the blue, although it is most likely
due to the bad weather (clouds) during the observations.
We note that besides the $g$-band, the flux densities derived from the spectroscopy
are consistent with the \hst\ and SDSS photometry, and that
the error in the absolute flux level in the blue part of the spectrum of Source 30
does not affect our analysis or conclusions.

\subsubsection{H$\alpha$ Emission}\label{sec_ha}

% Ha
\begin{deluxetable*}{ccccccc}
\tabletypesize{\footnotesize}
\tablecolumns{7} 
\tablewidth{0pt} 
\tablecaption{\ha\ Emission and Derived Properties of the YMCs\label{tab_ha}} 
\tablehead{ 
\colhead{ }  &  \colhead{Observed Flux} & \colhead{Extinction Corrected Flux} & \colhead{Equivalent} &
\colhead{N$_{\rm Lyc}$} & \colhead{Age} & \colhead{Mass} \\
\colhead{ }  & \colhead{(\fluxunits)} & \colhead{(\fluxunits)} & \colhead{Width (\AA)} & \colhead{($10^{49}$ s$^{-1}$)} &
\colhead{(Myr)} & \colhead{($10^3$ M$_\odot$)} }
\startdata
Source 26 & & & & & & \\
Photometry    &  $1.1(0.1) \times 10^{-12}$ & $1.6(0.2) \times 10^{-12}$ & 1600(370) & 230(40) & (0.1)2.9(0.8) & 65(12) \\
Spectroscopy  &  $9.5(0.3) \times 10^{-13}$ & $1.4(0.1) \times 10^{-12}$ & 1500(50) & 200(30) & (0.1)3.0(0.1) & 62(9) \\
 & & & & & & \\ 
Source 30 & & & & & & \\
Photometry    &  $7.4(0.9) \times 10^{-14}$ & $1.2(0.1) \times 10^{-13}$ & 3160(730) & 17(3) & $\lesssim 2.6$ & 4(1) \\
Spectroscopy  &  $7.3(0.2) \times 10^{-14}$ & $1.2(0.1) \times 10^{-13}$ & 3260(120) & 17(3) & $\lesssim 0.4$ & 4(1) \\
\enddata
\tablecomments{Uncertainties are shown in parentheses.}
\end{deluxetable*} 

In studies of extragalactic massive star forming regions, narrow-band imaging is commonly used to
estimate the emission of \ha, which in turn is used to infer physical properties such as
ionizing luminosity, age and mass.  Since we have spectroscopy and narrow-band imaging of our
target clusters, we compare the flux and equivalent width of \ha\ from both data sets to 
see how well the measurements agree.  

Using the \hst\ photometry, the \ha\ flux, $F_{{\rm H}\alpha}$, and equivalent width,
$W_{{\rm H}\alpha}$, are given by

\begin{eqnarray}
\label{haflux} F_{{\rm H}\alpha} = \left[\left(f_{\rm F658N} - f_{\rm cont}^{({\rm H}\alpha)}\right)
\Delta\lambda_{\rm F658N}\right] - \\
\left[\left(f_{\rm F660N} - f_{\rm cont}^{({\rm NII})}\right) \Delta\lambda_{\rm F660N}\right],
\nonumber
\end{eqnarray}

\noindent
and

\begin{equation}
W_{{\rm H}\alpha} = {F_{{\rm H}\alpha} \over f_{\rm cont}^{({\rm H}\alpha)}},
\label{haew}
\end{equation}

\noindent
where $F_{{\rm H}\alpha}$ has units of erg~s$^{-1}$~cm$^{-2}$, $W_{{\rm H}\alpha}$ has
units of \AA\ and $f_{\rm cont}^{({\rm H}\alpha)}$
is the continuum flux density at 6563~\AA\ in erg~s$^{-1}$~cm$^{-2}$~\AA$^{-1}$.

In Equation \ref{haflux}, the first term in square brackets is the total H$\alpha$ + \NII\ $\lambda6584$ flux
and the second term in square brackets is the total \NII\ $\lambda6584$ flux.\footnote{The \NII\
line at 6548~\AA\ is a relatively weak contaminant to the \ha\ flux and the filter throughput
drops at this wavelength.}
The measured flux densities and widths of the F658N and F660N filters are given by  
$f_{\rm F658N}$, $f_{\rm F660N}$ and $\Delta\lambda_{\rm F658N}$, $\Delta\lambda_{\rm F660N}$, respectively.
The flux densities of the continuum values, $f_{\rm cont}^{({\rm H} \alpha)}$ and $f_{\rm cont}^{({\rm NII})}$,
are found by interpolating between the F550M and $i$-band data points.

Table \ref{tab_ha} lists the measured \ha\ fluxes and equivalent widths obtained
from the photometry and from direct measurements of the \ha\ line in the spectra
using the IRAF task \texttt{splot}.  These values agree within the errors,
confirming that the appropriate imaging data can be used to obtain reliable \ha\
measurements.  

Table \ref{tab_ha} also lists extinction corrected \ha\ fluxes using internal extinctions of
\av=0.4 for Source 26 and \av=0.5 for Source 30 (see \S\ref{sec_sb99}).  Galactic foreground
extinction is also accounted for \citep[$E(B-V)=0.019$,][]{Schlegel98}.
A comparison of the \ha\ measurements with radio continuum observations \citep{Reines08a}
suggest that these \av\ may actually underestimate the true extinctions
of these infant clusters since optical observations may not probe
the deepest embedded regions of the YMCs.  Nonetheless, we use the optically derived
extinctions here.

\section{PHYSICAL PROPERTIES OF THE CLUSTERS DERIVED FROM H$\alpha$ EMISSION}\label{sec_lines}

\subsection{Ionizing Luminosities}

The flux and equivalent width of the \ha\ emission line can be used
to infer properties of a YMC such as the number of ionizing photons produced by the
massive stars, the cluster's age, and its mass \citep[e.g.][]{Alonso96,Reines08b}.  
We calculate ionizing luminosities, N$_{\rm Lyc}$, using the extinction corrected
\ha\ luminosities (Table \ref{tab_ha}) in the following equation derived from
\citet{Condon92} for a $10^4$ K gas:

\begin{equation}
\left({N_{\rm Lyc} \over {\rm s^{-1}}}\right) \gtrsim 7.87\times10^{11}
\left({L_{{\rm H}\alpha} \over {\rm erg ~s^{-1}}}\right).
\label{Nlyc}
\end{equation}

\noindent
Adopting a distance of 3.9 Mpc to \ngc,
the production rates of ionizing photons are approximately $2 \times 10^{51}$ and
$1.7 \times 10^{50}$ s$^{-1}$, or the equivalent of $\sim$200 and $\sim$17 O7.5 V
stars \citep{Vacca96} for Sources 26 and 30, respectively.  We note that the
ionizing luminosity of Source 30 derived here is lower than that in \citet{Reines08a}
($2.7 \times 10^{50}$ s$^{-1}$).  In that paper, the ionizing luminosity was derived
from radio continuum measurements which probe higher extinction regions than the \ha\
line used here.

\subsection{Ages}

\begin{figure}
\begin{center}
\includegraphics[scale=0.5]{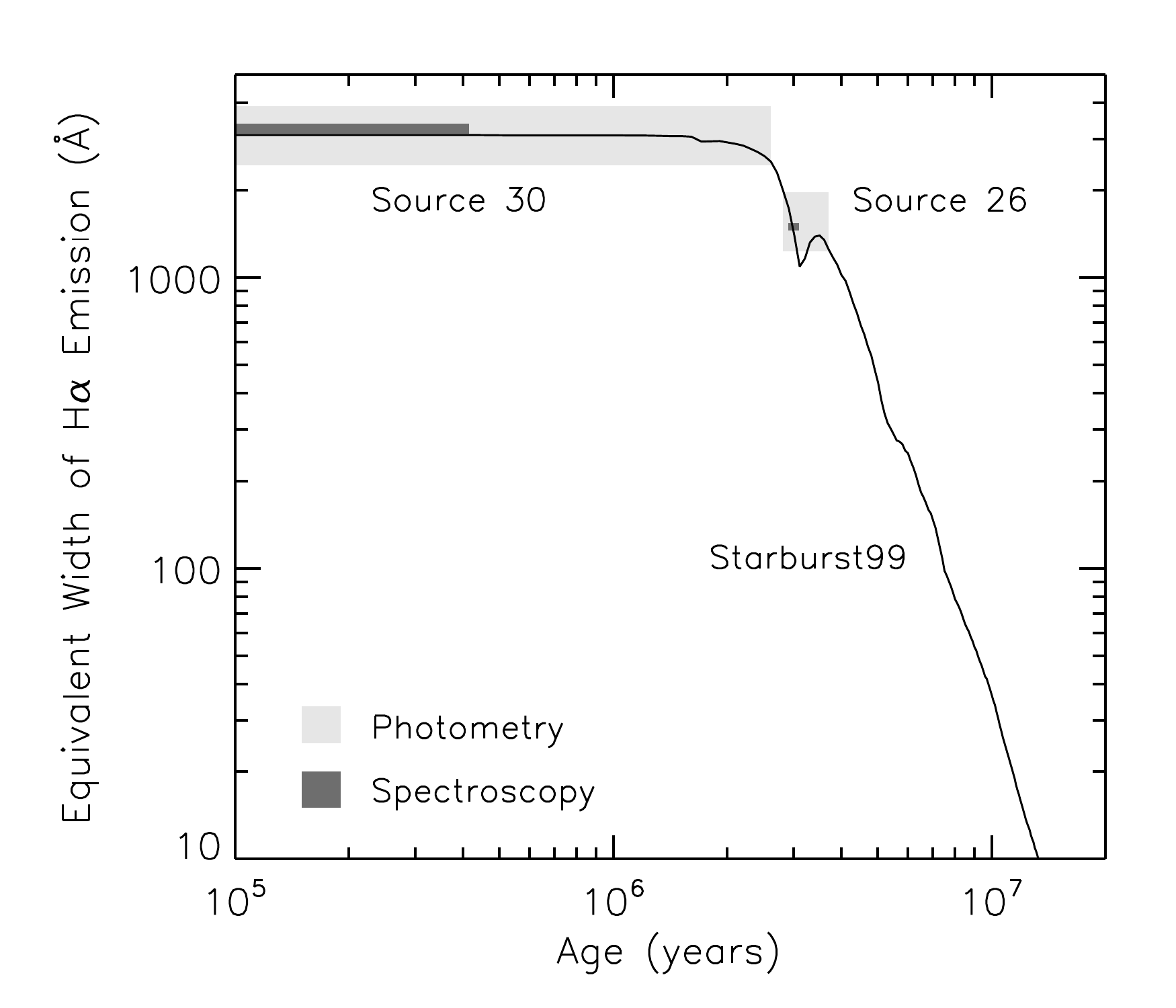}
\caption{Starburst99 model evolutionary track for the equivalent width of
\ha\ emission (see \S\ref{sec_sb99} for a description of the model).
The measured \ha\ equivalent widths of Sources 26 and 30, from both the
photometry and spectroscopy, are shown indicating their young
ages.  Uncertainties are reflected in the sizes of the shaded regions.}\label{plot_haew}
\end{center}
\end{figure}

Ages of the YMCs are estimated from their \ha\ equivalent widths.  
The equivalent widths from both the photometry and spectroscopy
are plotted in Figure \ref{plot_haew} with a
Starburst99 model as a function of age.
We note that the model equivalent widths
include a nebular component in the continuum (as do our measurements).
The \ha\ equivalent width of Source 26 suggests the cluster is $\approx$ 3 Myr old.
Source 30 has an \ha\ equivalent width consistent with the
maximum value, and therefore the youngest ages, of the model evolutionary track.
Source 30 is certainly younger than Source 26 and is likely $\lesssim 1$ Myr old.

\subsection{Masses}

\begin{figure*}
\begin{center}$
\begin{array}{cccc}
{\includegraphics[width=3.3in]{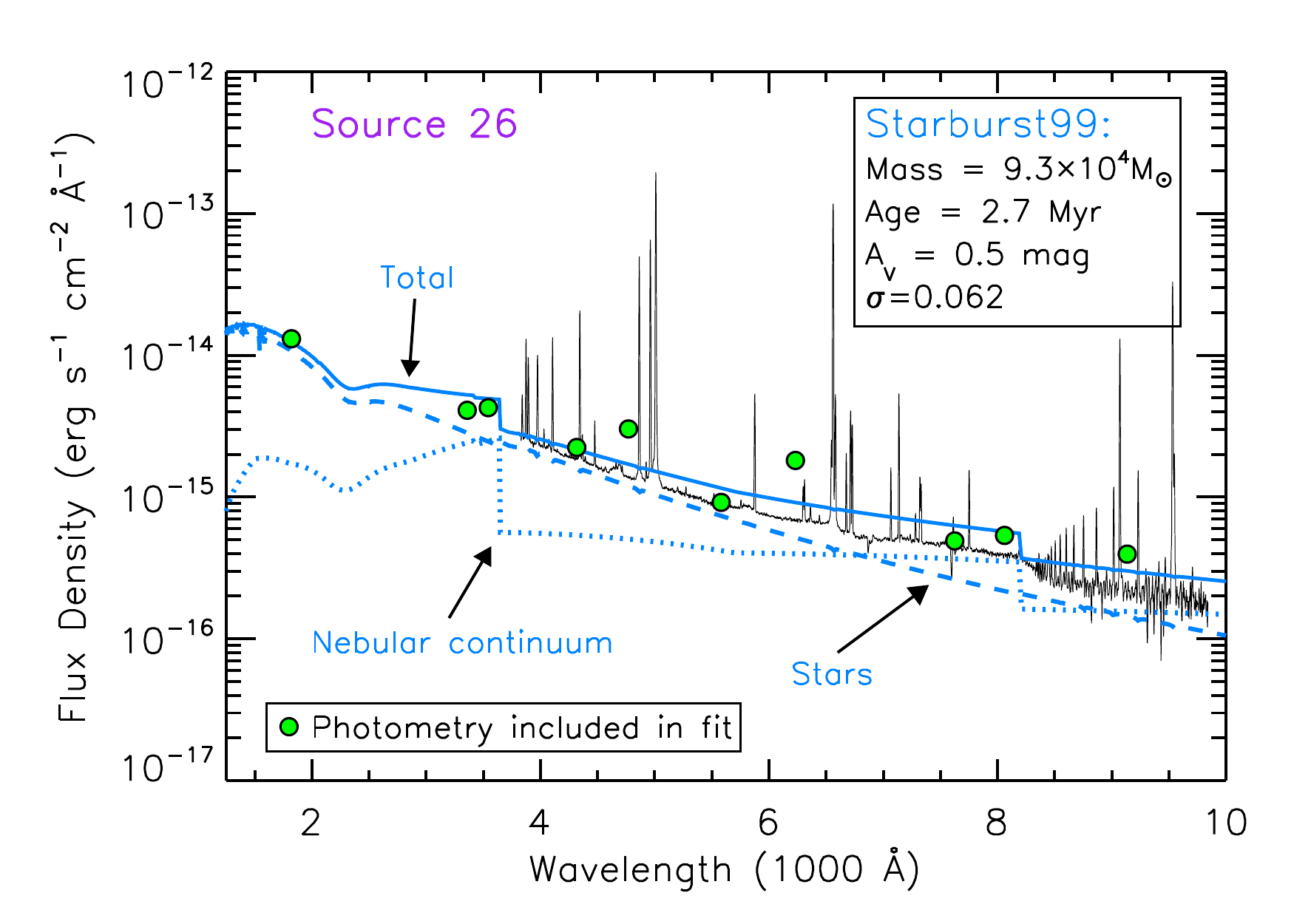}} &
{\includegraphics[width=3.3in]{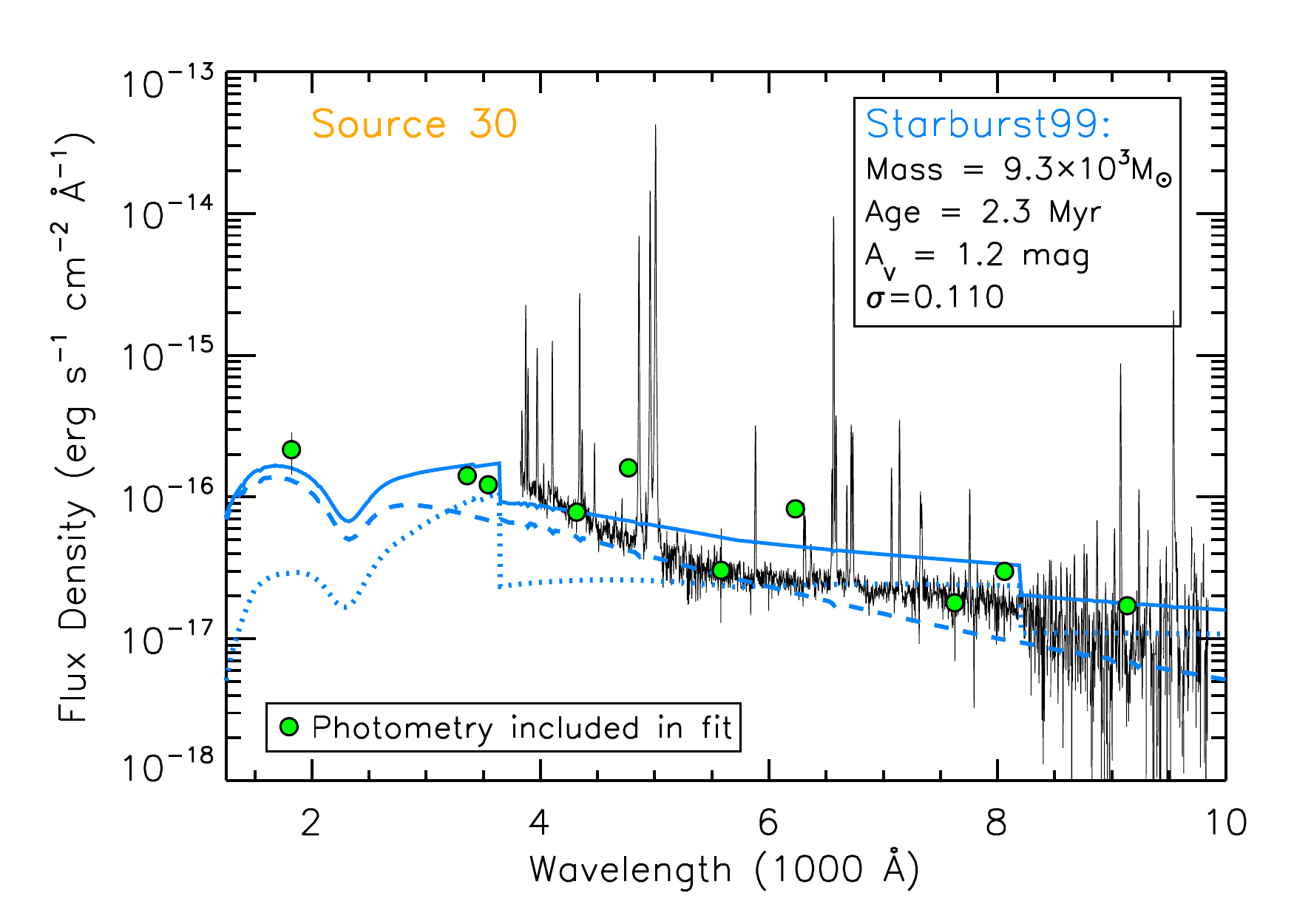}} \\
{\includegraphics[width=3.3in]{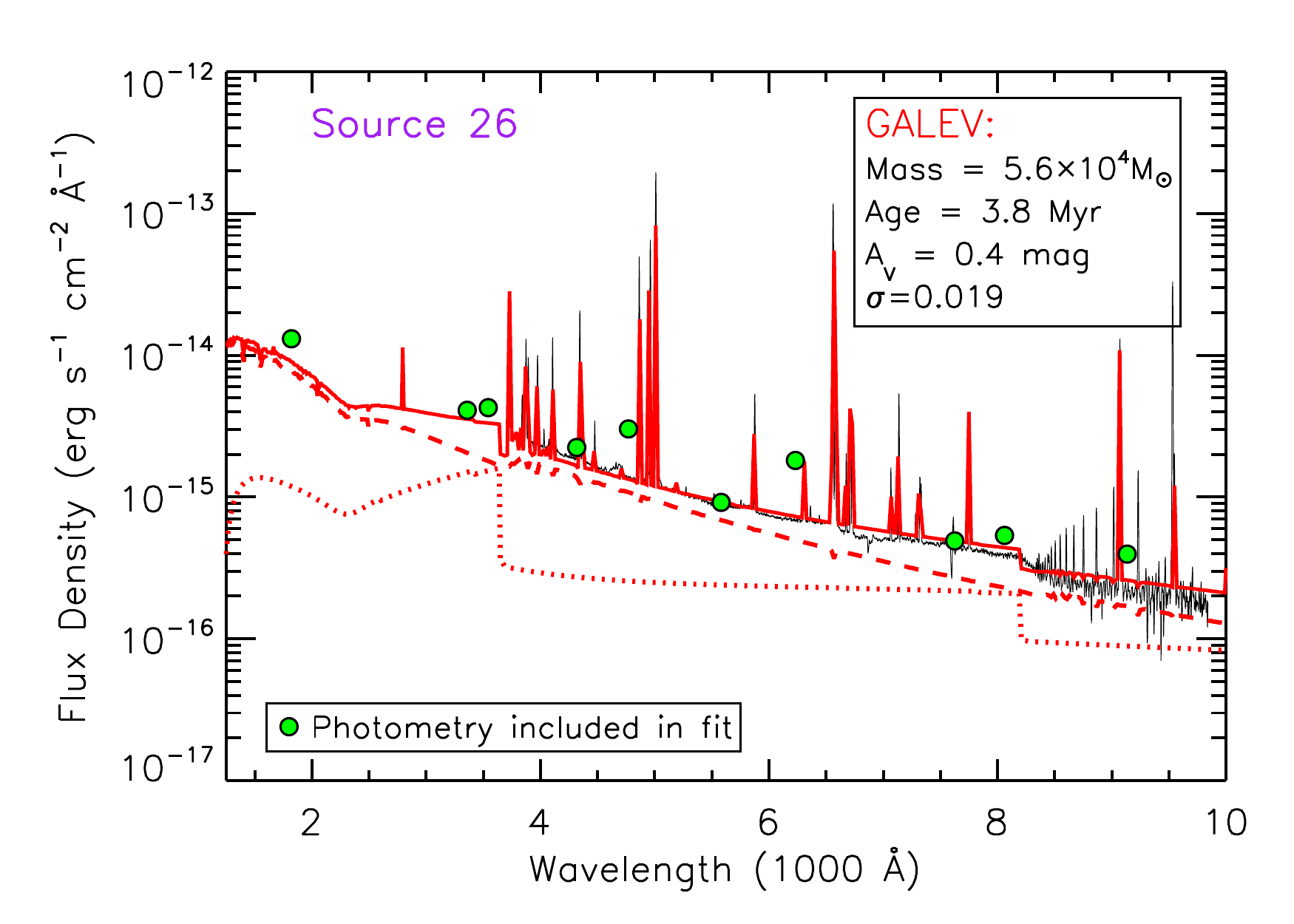}} &
{\includegraphics[width=3.3in]{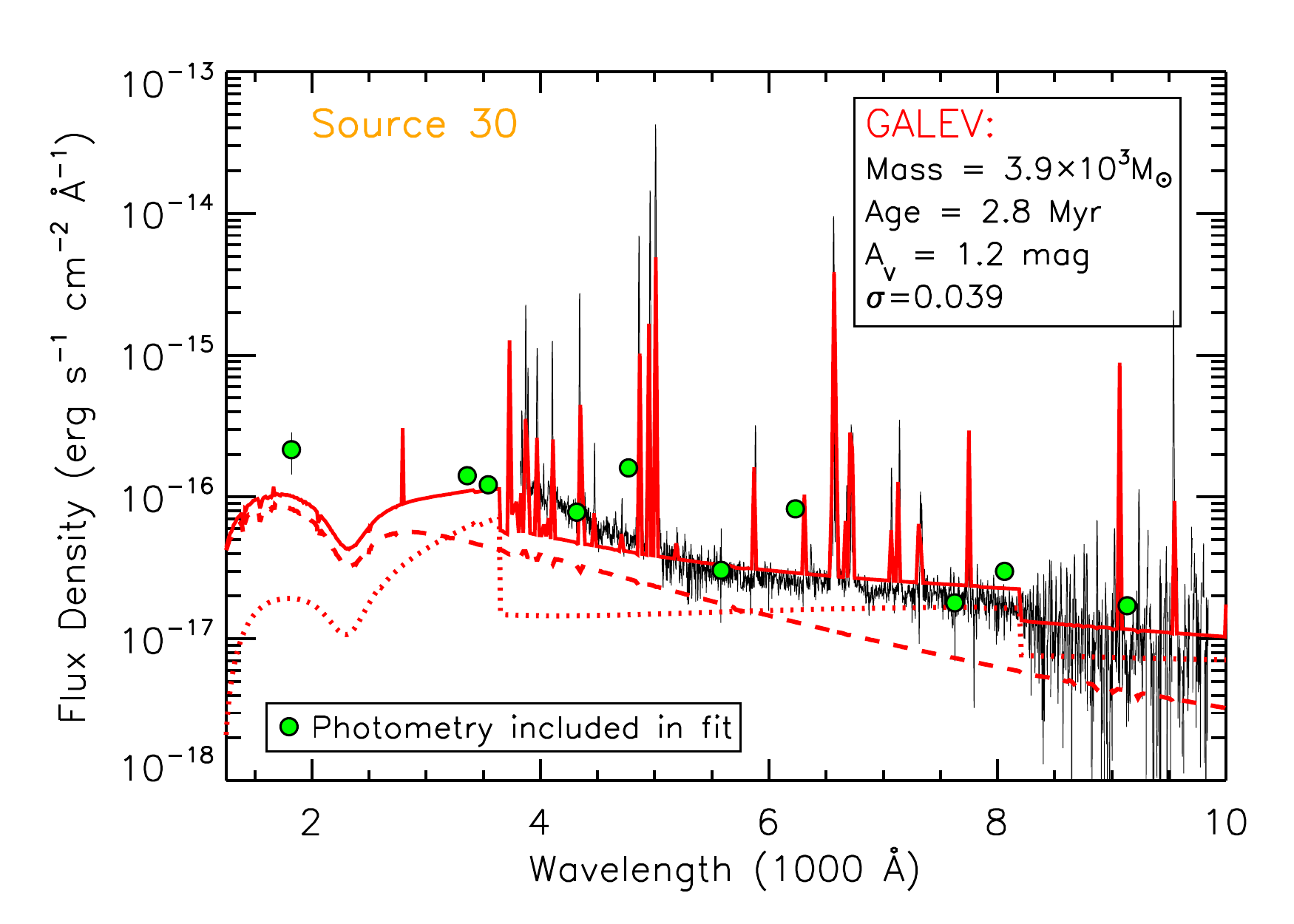}}
\end{array}$
\end{center}
\caption{Best-fitting Starburst99 (top) and GALEV (bottom) model SEDs for Sources
26 (left) and 30 (right) using all of the photometric bands (green dots) in the
fitting process.  Both models include stellar continuum (dashed lines) and nebular
continuum (dotted lines).  The GALEV models also include emission lines.  Model
parameters (masses, ages and extinctions) are shown in the upper right corners of
the plots.  The goodness-of-fit parameter, $\sigma$, is also shown for each model.
A lower value of $\sigma$ indicates a better fit.
\label{plot_seds_all}}
\end{figure*}

Assuming mass scales with ionizing luminosity, masses of the clusters can be estimated by
multiplying an input model mass by the ratios of the measured ionizing luminosities to
the model ionizing luminosities at their respective ages.  Using the Starburst99 model described in
\S\ref{sec_sb99}, we infer
masses of $\sim 6 \times 10^4$ and $\sim 4 \times 10^3$ M$_\odot$ for Sources 26 and 30,
respectively.  Again, the mass of Source 30 is lower than that presented in \citet{Reines08a}
($\sim 6 \times 10^3$ M$_\odot$) which used the radio derived ionizing luminosity.
Ionizing luminosities, ages, and masses of Sources 26 and 30, based on the \ha\ measurements
presented in \S\ref{sec_ha}, are summarized in Table \ref{tab_ha}.
It is noted that caution must be applied when interpreting the physical
properties of observed clusters derived from population synthesis models,
since an underlying assumption of every Simple Stellar Population (SSP) model
is a well-populated Initial Mass Function (IMF) and stochastic fluctuations can
bias the results.  

\section{MODELING THE STARBURST SPECTRAL ENERGY DISTRIBUTIONS}\label{sec_model}

In addition to using hydrogen recombination lines such as \ha,
broad-band photometric observations of extragalactic massive star clusters are commonly
compared to models of SSPs in order to estimate their physical properties such as age,
extinction and mass.  Here we perform SED fitting to our {\it photometric} data using the
Starburst99 \citep{Leitherer99} and GALEV \citep{Kotulla09} evolutionary synthesis models,
and investigate how well the model SEDs fit the continua in our observed {\it spectra}.
In addition to the stellar light, both sets of models include nebular continuum (although these
are computed in slightly different ways).  The GALEV models also have the option to include a set of
metallicity-dependent emission lines \citep{Anders03}.

Physical properties of the clusters are estimated by comparing a grid of model
SEDs (Starburst99 and GALEV separately) to measured photometric flux densities.  The grid
includes SEDs with ages $< 50$ Myr in steps of 0.1 Myr, and visual extinctions $< 3$ in
increments of 0.1 magnitudes.  We apply a 30 Doradus extinction curve
adopted from \citet[][Table~3]{Misselt99} and \citet[][Table~6]{Fitzpatrick85}, using
the parameterization given by \citet{Fitzpatrick90}.  Galactic foreground extinction towards \ngc\
is also accounted for \citep[$E(B-V)=0.019$,][]{Schlegel98} using the extinction curve
of \citet{Cardelli89}.  Each model SED, of a given age and \av, is then convolved with
the total system throughput curves for the photometric filters to obtain synthetic
flux densities.  The best-fit model SED is determined by minimizing a goodness-of-fit
parameter, $\sigma$, equal to the standard deviation of the logarithmic residuals
of the observed and synthetic flux densities (weighted by the errors of the observed
flux densities).  A mass estimate is obtained by scaling the model mass by the mean
logarithmic offset between the observed and best-fit model flux densities.

In all, we ran our SED fitting routine four times for each cluster: 1) using
Starburst99 and all available broad-band photometric data, 2) using Starburst99 and only
photometric data with minimal line emission, 3) using GALEV (with emission
lines) and all available photometric data, and 4) using GALEV (without emission lines)
and only photometric data with minimal line emission.  Stellar and nebular continuum
emission is included in the models for each case.  Filters with minimal
line emission were identified using our Figure \ref{plot_spectra} and Table 1
in \citet{Anders03} (for the short-wavelength filters not covered by our spectra).
These include the F170W, F336W, $u$, F550M, and $i$ filters.  Details about the
models and the results of SED fitting are described below.

\subsection{Starburst99: Stellar and Nebular Continuum}\label{sec_sb99}

We use the latest Starburst99 models (Version 5.1) with a
metallicity of $Z=0.004$ \citep[as appropriate for NGC~4449,][]{Lequeux79},
an instantaneous burst of $10^4$ M$_\odot$ with a Kroupa IMF ($0.1-100$ M$_\odot$), the
Geneva evolutionary tracks with high mass loss, and the Pauldrach/Hillier atmospheres.
The models include stellar and nebular continuum but no emission lines.

SED fitting is first performed using all of the available broad-band photometric data
in this study (F170W, F336W, $u$,
F435W, $g$, F550M, $r$, $i$, F814W, and $z$).  The best-fitting model SEDs for Sources 26 and 30 are
shown at the top of Figure \ref{plot_seds_all}.  It is clear that the model SEDs do a poor
job fitting the spectral continua.  This is because the spectra contain strong emission lines which
are not accounted for in the models.  The photometric bands containing significant line emission
cause the model SEDs to be higher than the observed continua.

\begin{figure*}
\begin{center}$
\begin{array}{cccc}
{\includegraphics[width=3.3in]{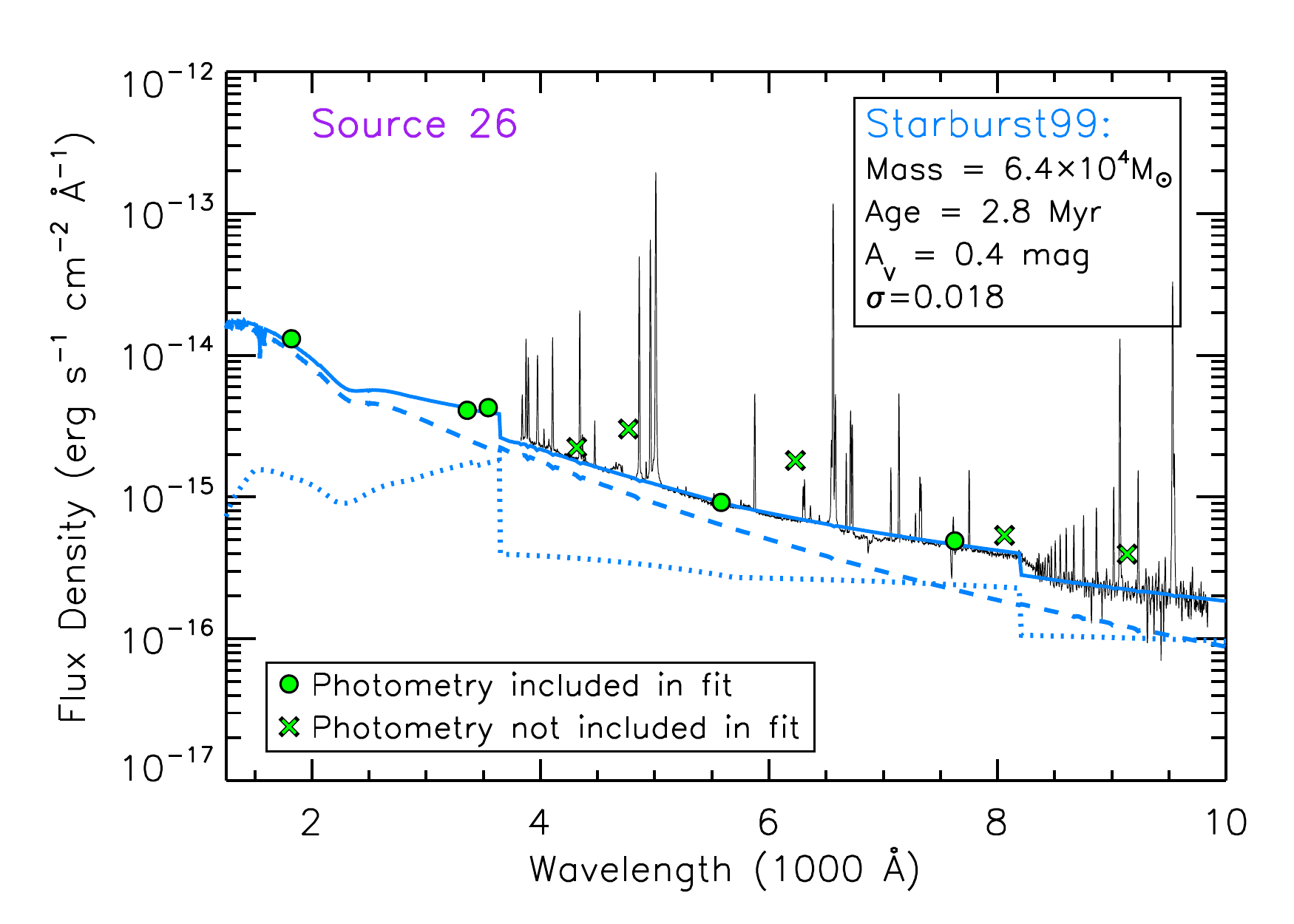}} &
{\includegraphics[width=3.3in]{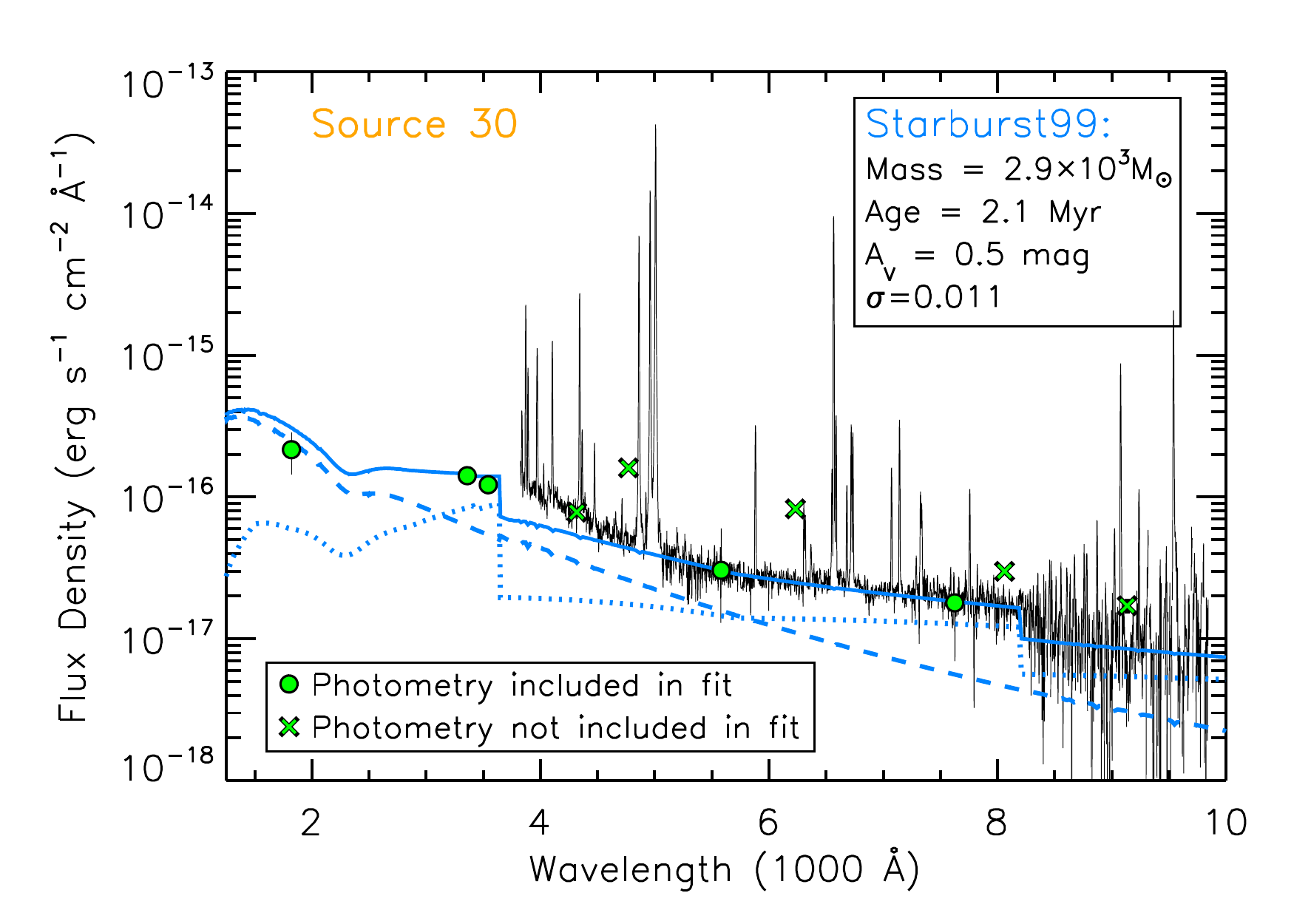}} \\
{\includegraphics[width=3.3in]{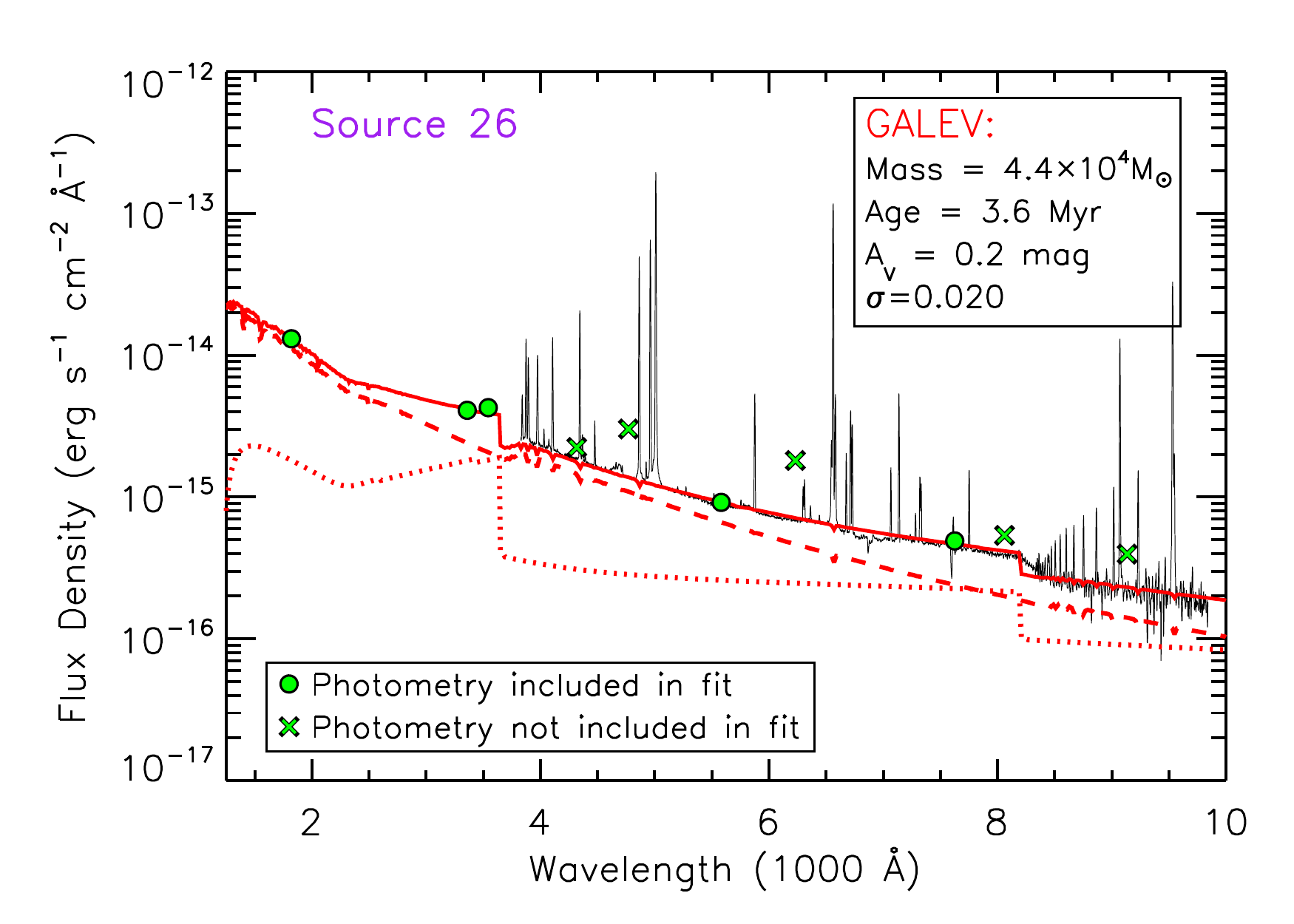}} &
{\includegraphics[width=3.3in]{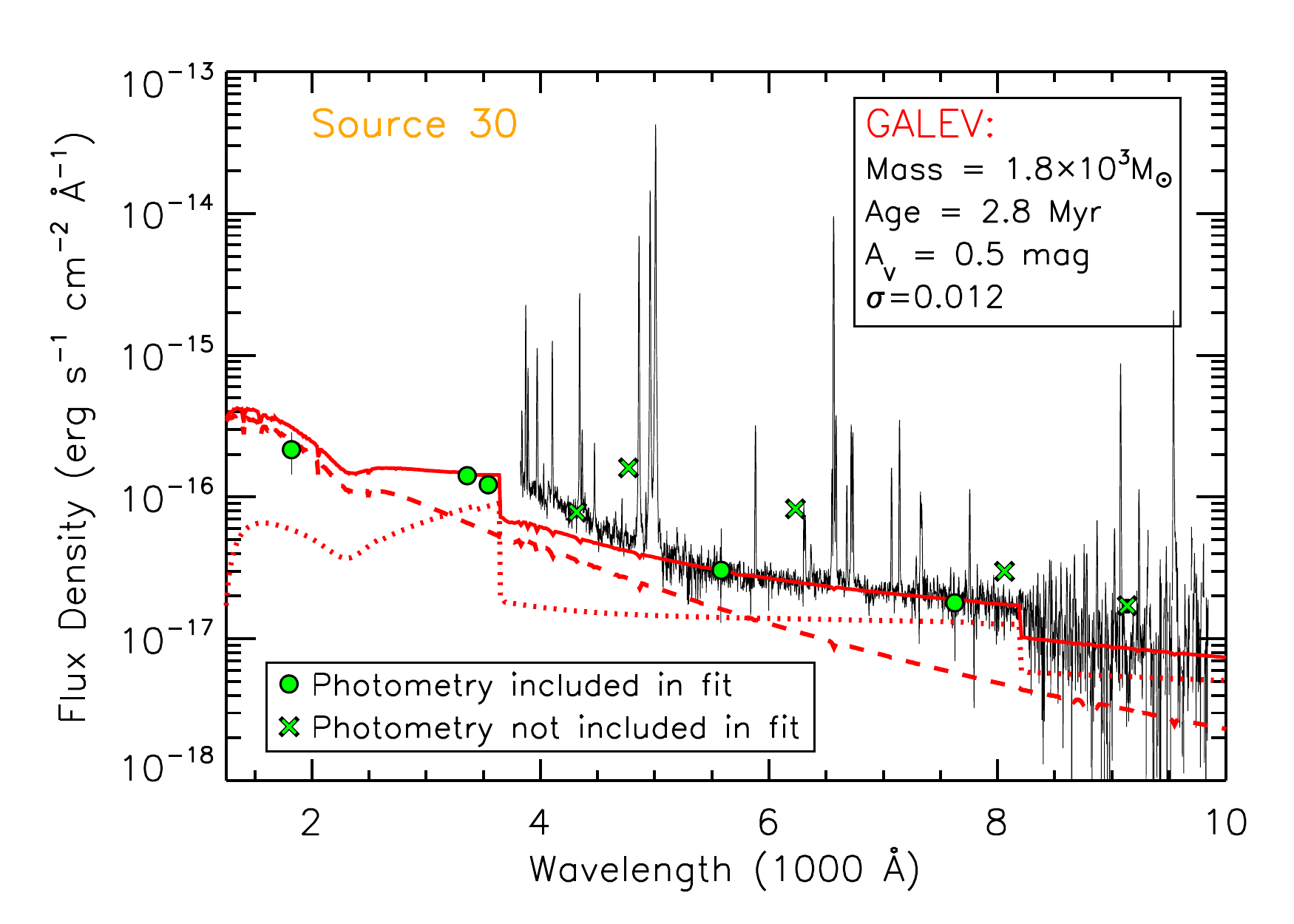}}
\end{array}$
\end{center}
\caption{Same as Figure \ref{plot_seds_all} except that only photometric bands
with minimal line emission are included in the fitting process.  Also,
emission lines are not included in the GALEV models used here. 
\label{plot_seds}}
\end{figure*}

Next, we perform SED fitting using only those filters with minimal line emission
(F170W, F336W, $u$, F550M, and $i$).  In this case, the best-fit models do a very good job fitting
both the photometry included in the fitting process and the observed
continua of the spectra\footnote{The flux calibration of the blue end of the
observed spectrum of Source 30 is in error (see \S\ref{sec_fluxden}).  However, this
does not affect our results since the photometry is being fit.} (see Figure \ref{plot_seds}).
Also, the inferred physical properties of the clusters (shown in Figure \ref{plot_seds}) are
consistent with those derived from the \ha\ emission (see Table \ref{tab_ha}). 

Comparing the best-fit Starburst99 model parameters from the two fitting processes
(all bands versus bands with minimal line emission), we find that mass estimates can
vary significantly.  For Source 30, the mass derived from fitting all of the photometry
is a factor of 3 times too high due to strong lines raising the overall SED.  The extinction
estimate for Source 30 is also higher by 0.7 mag when all of the bands are fit since 
the model SED needs to be flatter in the red to better match the F814W and $z$-band data
(which contain significant line emission).

The impact of nebular continuum on the total SEDs is striking in both clusters
(see Figure \ref{plot_seds}), especially in the $U$- and $I$-band
regions of the spectra where the Balmer (3646 \AA) and Paschen (8207 \AA) discontinuities
are located.  These discontinuities, or ``jumps'', in the nebular continuum emission 
are due to hydrogen and helium free-bound recombination processes.
The nebular continuum actually dominates the total continuum in these
wavelength regions for the younger YMC, Source 30.  

\begin{figure}
\begin{center}
\includegraphics[scale=0.65]{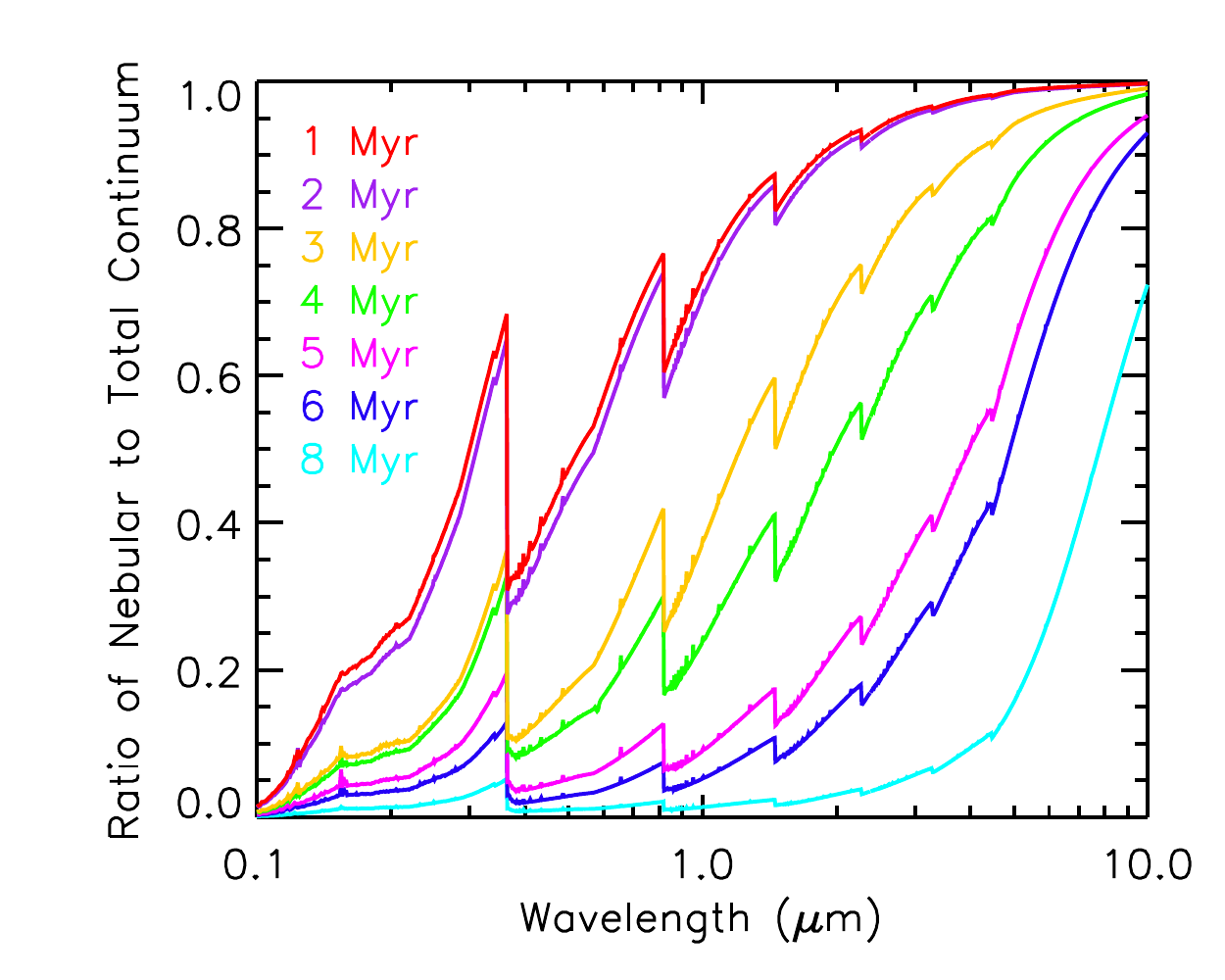}
\caption{Ratio of the nebular continuum to the total continuum as a function of
wavelength for a Starburst99 SSP (\S\ref{sec_sb99}) at various ages.
At young ages, nebular continuum is a significant component
of the total continuum at optical wavelengths, especially shortward of the
Balmer (3646 \AA) and Paschen jumps (8207 \AA) located in the
$U$- and $I$-band wavelength regions.  The impact of emission lines is
not included here.}\label{plot_gasfrac}
\end{center}
\end{figure}

To further illustrate the importance
of nebular continuum in young stellar populations, we plot the ratio
of nebular continuum to the total continuum as a function of wavelength for
Starburst99 model SEDs ($Z=0.004$) at various ages (Figure \ref{plot_gasfrac}).
It is clear that, at young ages, nebular continuum is a significant component
of the total continuum, even at optical wavelengths.

\subsection{GALEV: Continuum Plus Emission Lines}\label{sec_galev}

GALEV models were kindly provided to us by Ralf Kotulla and the GALEV team.
These models were run using the Geneva evolutionary tracks with a minimum age and
time resolution of 0.1 Myr, unlike the models available on the GALEV website which
use the Padova isochrones and have a minimum age and time step of 4 Myr.  The models
also use a ``fraction of visible mass'' equal to 1 rather than 0.5, which is used for the
standard models available on the web (Ralf Kotulla, private communication).  The models were run
with a metallicity of $Z=0.004$ and a Kroupa IMF ($0.1-100$ M$_\odot$).  

These models are very similar
to the Starburst99 models described above, with the notable exception that the GALEV models
include emission lines from nebular gas \citep{Anders03,Kotulla09}.  The flux of the hydrogen
lines are computed using atomic physics and the production rate of ionizing photons, whereas
non-hydrogen line strengths are computed using metallicity-dependent line ratios relative to
\hbeta.  In the case of low-metallicity gas (including $Z=0.004$), observed line ratios are
taken from the large sample of giant \HII\ regions and Blue Compact Dwarf (BCD)
galaxies presented in \citet{Izotov94,Izotov97} and \citet{Izotov98}.

We perform SED fitting with the GALEV models in the same way as the Starburst99 models.
First we fit all of the photometric filters and include emission lines in the models
(the option to include lines is not available for Starburst99).  The best-fitting model
SEDs for Sources 26 and 30 are shown in Figure \ref{plot_seds_all}.
It is clear that the GALEV models including emission lines do a much better job matching
the continua in the observed spectra than the Starburst99 models do when we fit all
of the photometry.  However, the GALEV model SEDs shown in Figure \ref{plot_seds_all} are still a
bit too high in the red to near-infrared and too low in the ultraviolet.
This is likely due to mismatched line intensities between the models and the observed spectra. 
We note that currently the GALEV models are missing the higher level Paschen lines, although they
do include Pa$\alpha$ through Pa$\epsilon$.
Overall, however, the GALEV models including emission lines perform much better than
the Starburst99 models when we fit all of the photometry including bands with significant line
contamination.

We also only fit the photometry with minimal line emission with GALEV SEDs as we did with the
Starburst99 models.  Here we exclude emission lines in the models so they only contain
components from the stellar and nebular continua.  The best-fitting SEDs are shown in Figure
\ref{plot_seds}.  These models perform equally well as the Starburst99 models fit only
to the photometry with minimal line emission (also shown in Figure \ref{plot_seds}).  However, it is
curious that the best-fit model parameters from the two families of models are as different
as they are.  The differences may be due to the ways in which the nebular continuum is calculated.
The best-fit GALEV SEDs shown in Figure \ref{plot_seds} also do a better job fitting the continua
of the observed spectra than the GALEV models (including lines) fit to all of the photometry
(Figure \ref{plot_seds_all}).

%Based on the results presented here, we recommend 
%using broad-band filters with minimal line emission
%if possible and 
	
\section{THE IMPORTANCE OF NEBULAR EMISSION IN BROAD-BAND PHOTOMETRY}\label{sec_broad-band}

A number of authors have discussed the impact of nebular emission on broad-band
photometry of young stellar populations based on synthesis modeling
\citep[e.g. ][]{Kruger95,Leitherer95,Zackrisson01,Anders03,Molla09}.  Our spectrophotometic
observations of infant massive star clusters clearly support the conclusion that nebular
emission, both in lines and continuum, significantly affects broad-band photometry of these objects.

\subsection{Contributions to Broad-band Fluxes}\label{sec_contributions}

In an effort to provide a benchmark for estimating the impact of nebular emission
on photometric observations of young massive stellar populations, we compute the relative contributions
of the stellar continuum, nebular continuum, and emission lines to the total flux of Source
26 through various \hst\ filter/instrument combinations (including WFC3) and the SDSS $griz$ filters.\footnote{We
do not use the spectrum of Source 30 since it has a low signal-to-noise
ratio and the flux calibration is uncertain in the blue (see \S\ref{sec_spec} and
\S\ref{sec_fluxden}).}
The contribution of nebular emission in the \hst/ACS WFC F814W filter is of
particular interest since, in their study of radio-detected clusters in \ngc,
\citet{Reines08a} found an excess in this band relative to model SEDs for a pure
stellar continuum.

\begin{figure}
\includegraphics[scale=0.65]{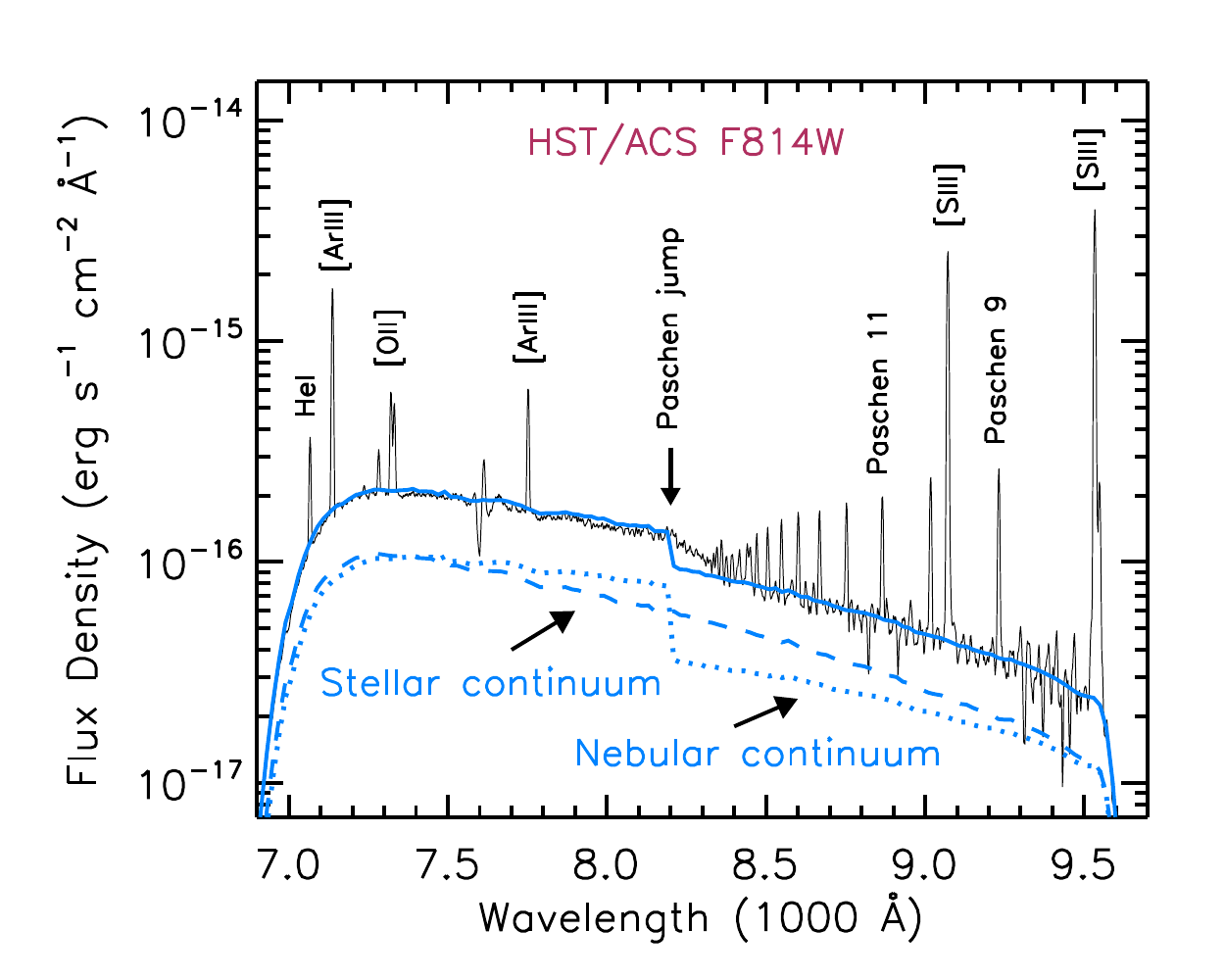}
\caption{Spectrum of Source 26 multiplied by the \hst/ACS WFC F814W total system throughput
curve.  The best-fitting Starburst99 model SED from Figure \ref{plot_seds} is shown as a solid
blue line.  The stellar and nebular continua are shown as dashed and dotted lines, respectively.
The relative contributions of the stellar continuum, nebular continuum and emission lines to the
total flux of this filter are approximately 40\%, 40\%  and 20\%.}\label{plot_f814w}
\end{figure}

The spectrum of Source 26 multiplied by the \hst/ACS WFC F814W total system throughput curve
is shown in Figure \ref{plot_f814w}.  The best-fitting Starburst99 model SED (from Figure \ref{plot_seds})
is also shown.  In addition to stellar and nebular continuum, the spectrum
contains several prominent emission lines in this band.  The strongest are \SIII\ $\lambda$9532 and
\SIII\ $\lambda$9069.  A number of other lines, including the Paschen series, are detected and
labeled in  Figure \ref{plot_f814w}.  

\begin{deluxetable*}{ccccc}
\tabletypesize{\footnotesize}
\tablecolumns{5}
\tablewidth{0pt}
\setlength{\tabcolsep}{0.07in}
\tablecaption{Relative Contributions to Broad-band Fluxes of Source 26\label{tab_contributions}}
\tablehead{
\colhead{Filter} & \colhead{Instrument} & \colhead{Stellar} & \colhead{Nebular} & \colhead{Emission} \\
\colhead{} & \colhead{} & \colhead{Continuum} & \colhead{Continuum} & \colhead{Lines} }
\startdata
\cutinhead{{\it HST} Medium and Wide Filters in the $I$-band Region}
F845M & WFC3/UVIS &  0.49 & 0.41 & 0.09 \\
F814W & WFC3/UVIS &  0.39 & 0.38 & 0.23 \\
F814W & ACS/WFC &  0.41 & 0.39 & 0.20 \\
F814W & ACS/HRC &  0.38 & 0.36 & 0.25 \\
F814W & WFPC2 &  0.41 & 0.40 & 0.20 \\
F791W & WFPC2 &  0.46 & 0.45 & 0.10 \\
F775W & WFC3/UVIS &  0.46 & 0.45 & 0.09 \\
F775W & ACS/WFC &  0.46 & 0.46 & 0.08 \\
F775W & ACS/HRC &  0.46 & 0.46 & 0.08 \\
F763M & WFC3/UVIS &  0.47 & 0.50 & 0.03 \\
\cutinhead{{\it HST} Medium and Wide Filters in the $V$-band Region}
F555W & WFC3/UVIS &  0.31 & 0.12 & 0.56 \\
F555W & ACS/WFC &  0.27 & 0.11 & 0.63 \\
F555W & ACS/HRC &  0.26 & 0.11 & 0.63 \\
F555W & WFPC2 &  0.32 & 0.13 & 0.54 \\
F550M & ACS/WFC &  0.70 & 0.31 & 0.00 \\
F550M & ACS/HRC &  0.70 & 0.31 & 0.00 \\
F547M & WFC3/UVIS &  0.71 & 0.30 & 0.00 \\
F547M & WFPC2 &  0.68 & 0.29 & 0.03 \\
\cutinhead{SDSS Filters}
$g$ & SDSS &  0.32 & 0.10 & 0.58 \\
$r$ & SDSS &  0.27 & 0.16 & 0.57 \\
$i$ & SDSS &  0.46 & 0.46 & 0.07 \\
$z$ & SDSS &  0.34 & 0.26 & 0.40 \\
\enddata
\end{deluxetable*}

\begin{figure*}
\begin{center}$
\begin{array}{cccccc}
{\includegraphics[width=3in]{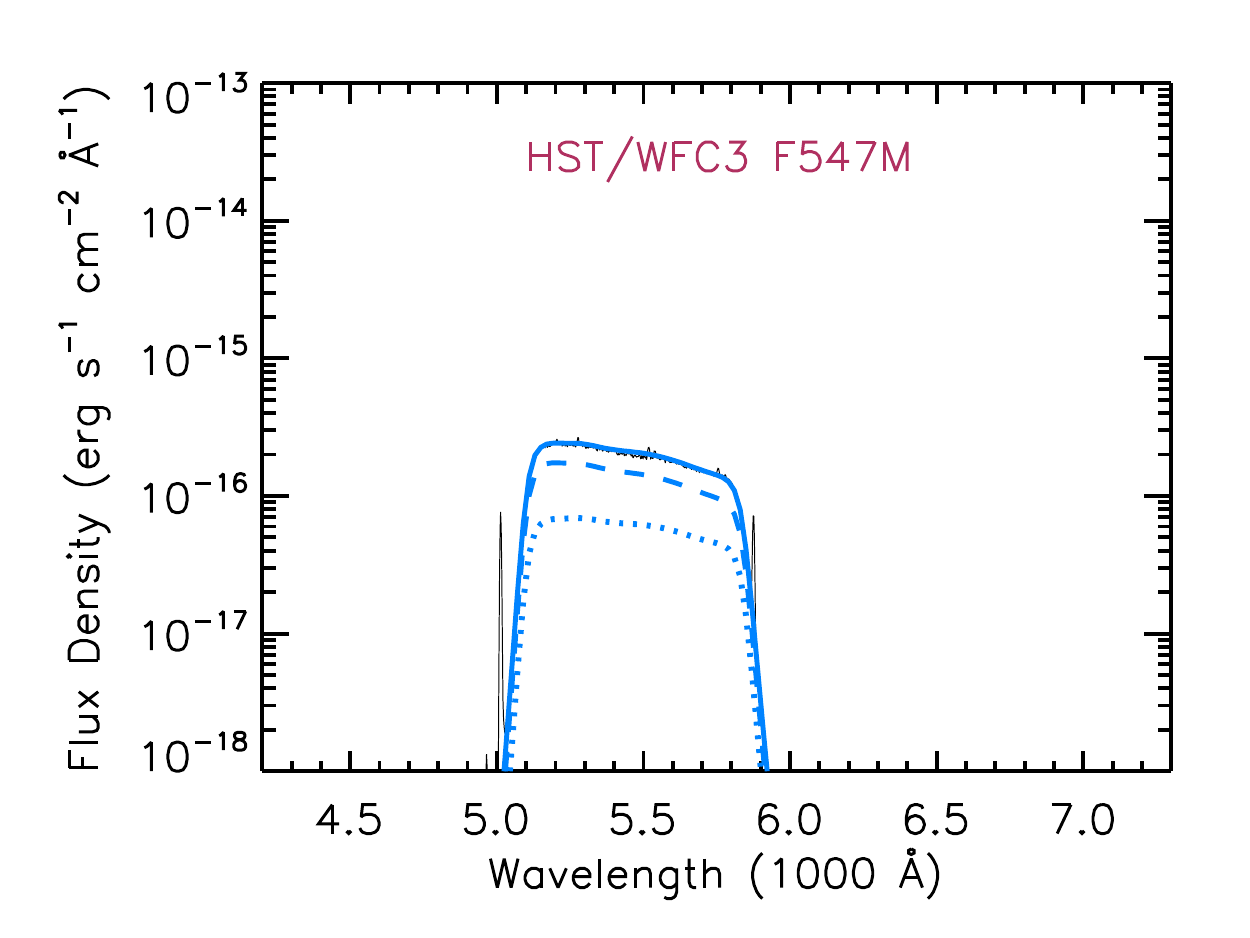}} &
{\includegraphics[width=3in]{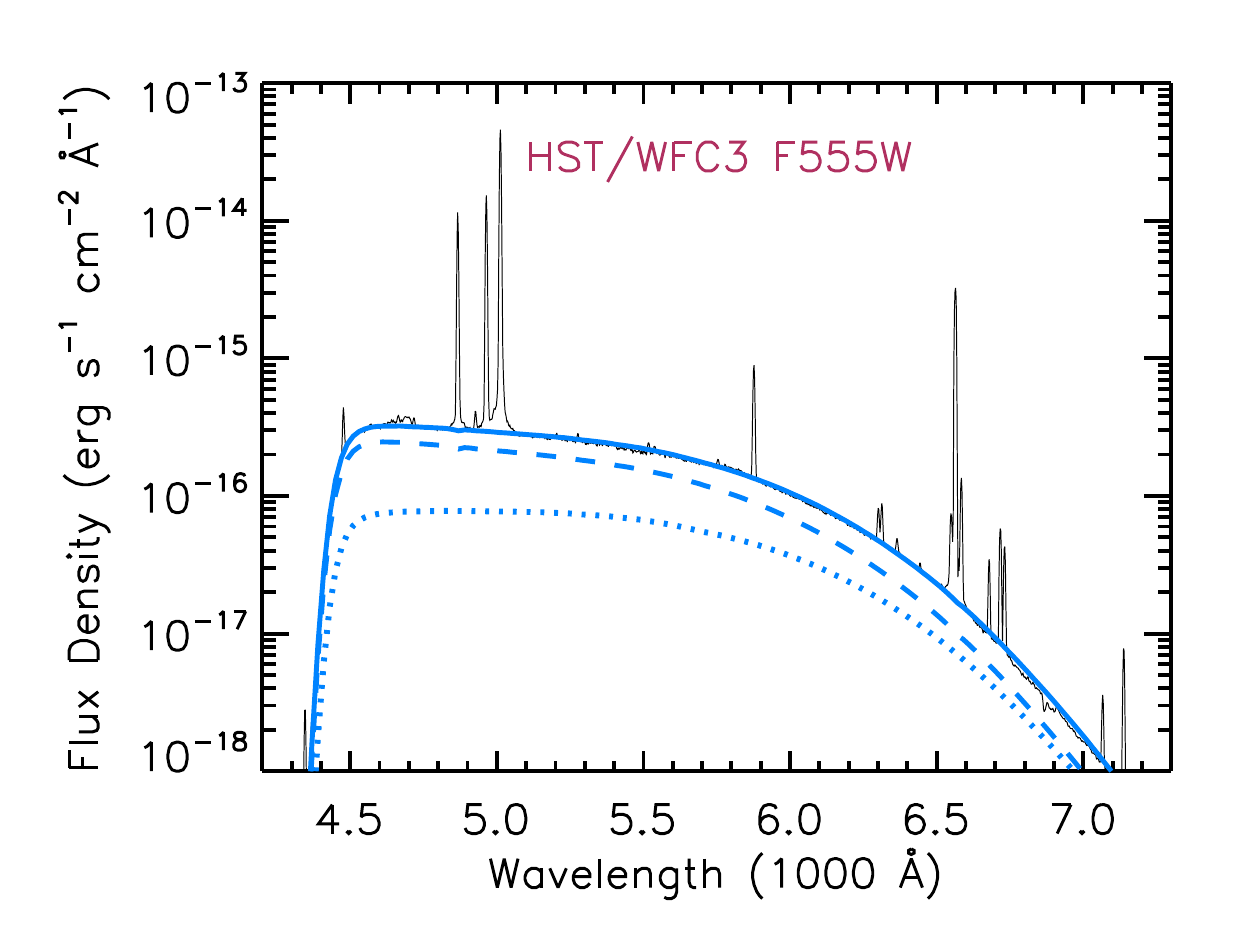}} \\
{\includegraphics[width=3in]{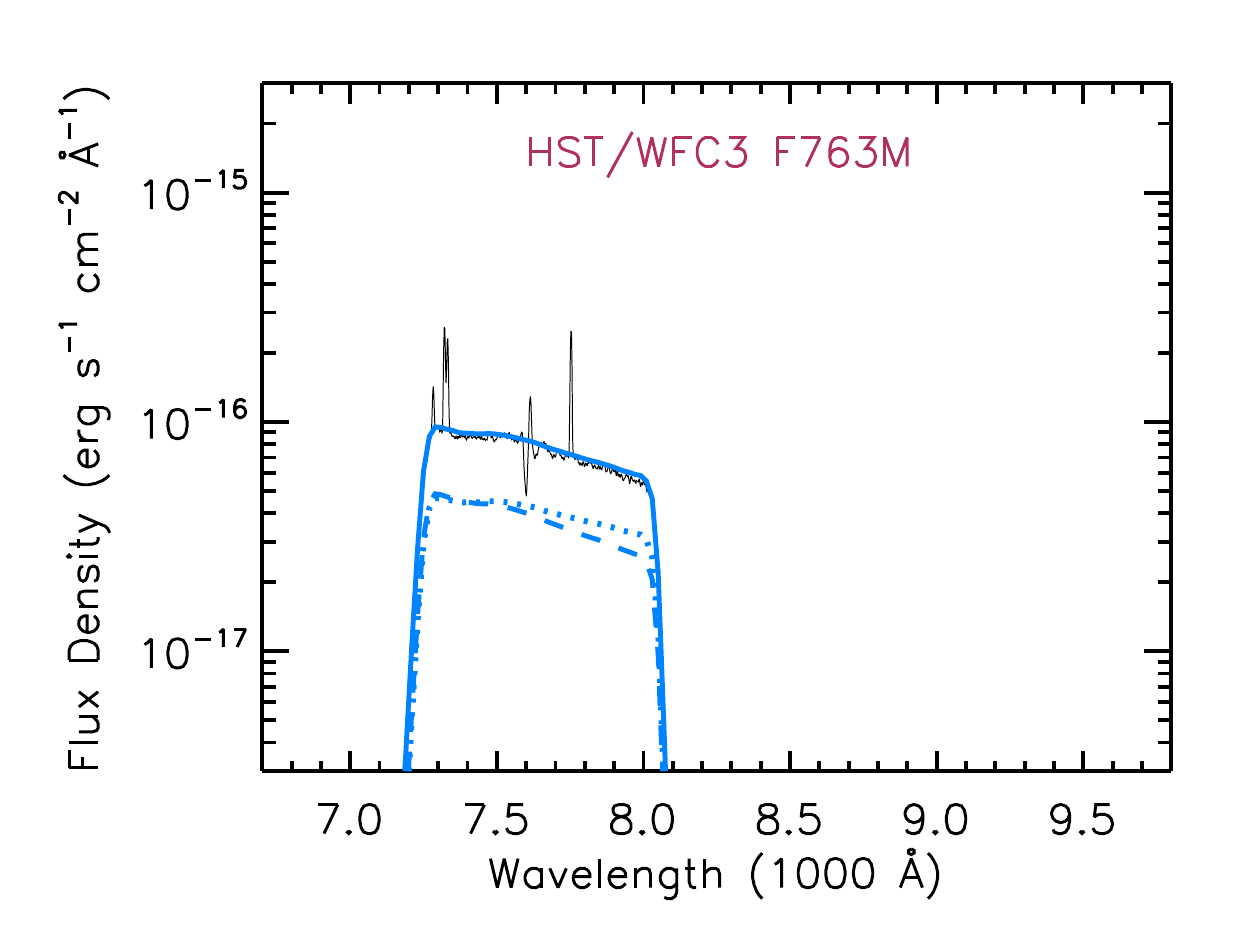}} &
{\includegraphics[width=3in]{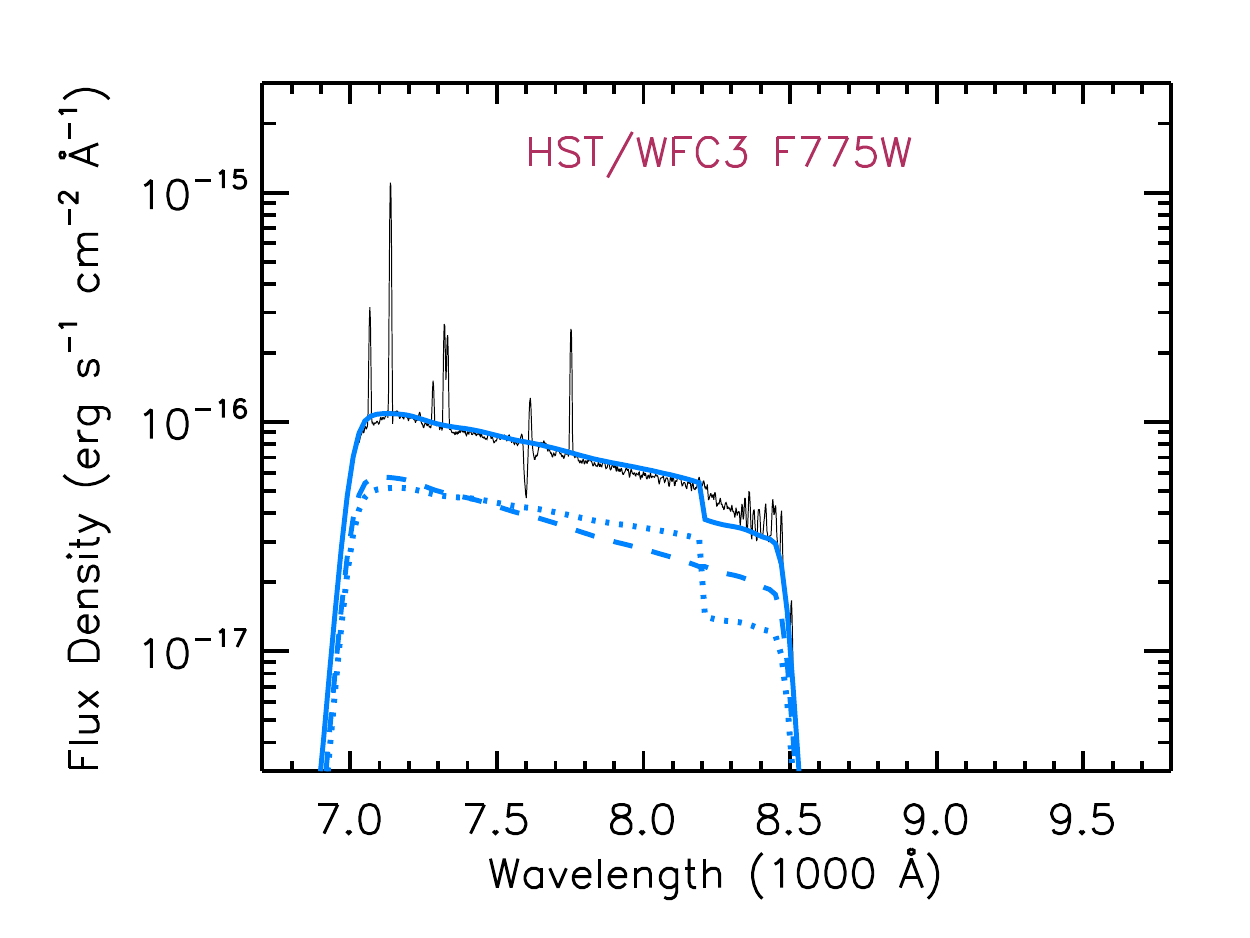}} \\ 
{\includegraphics[width=3in]{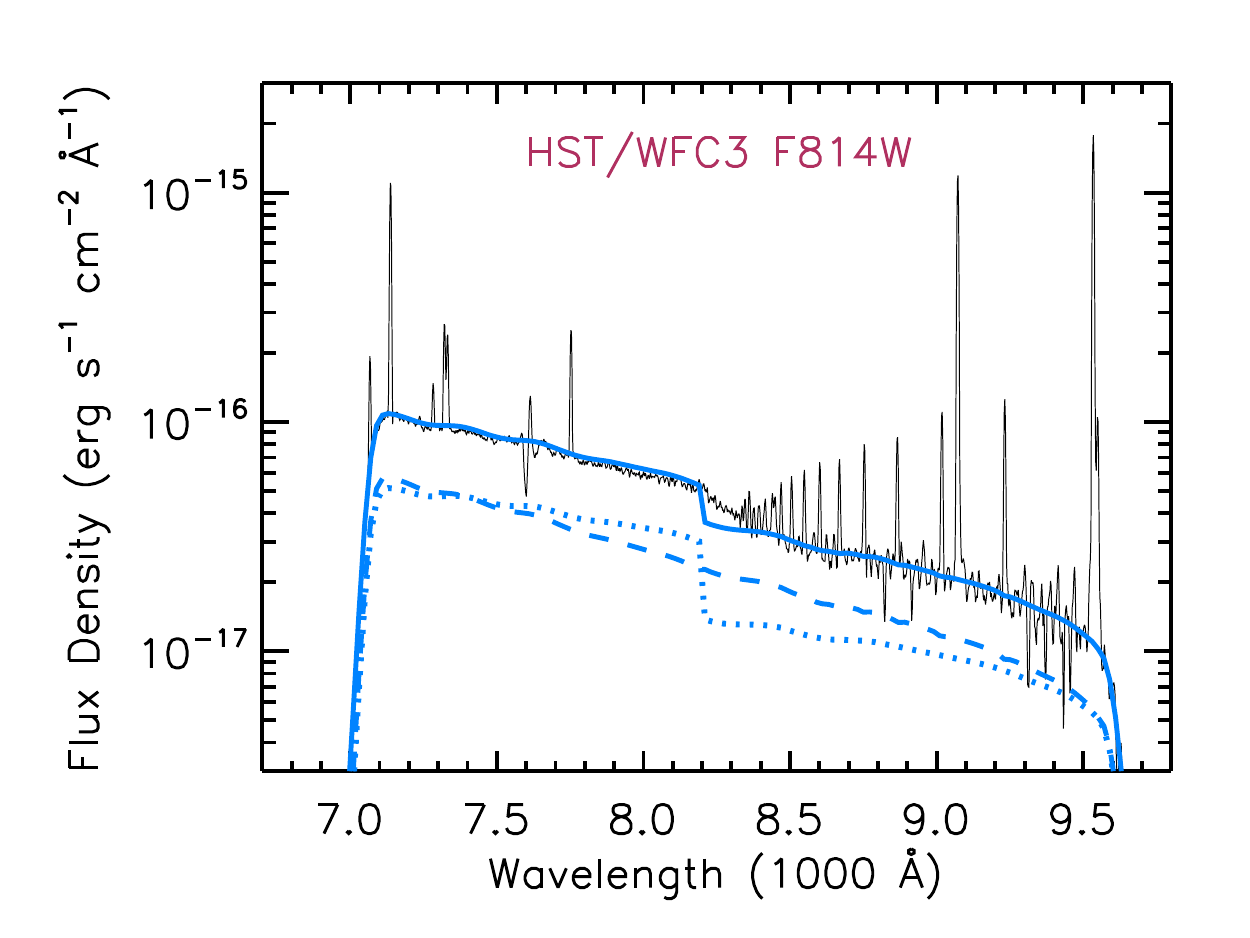}} &
{\includegraphics[width=3in]{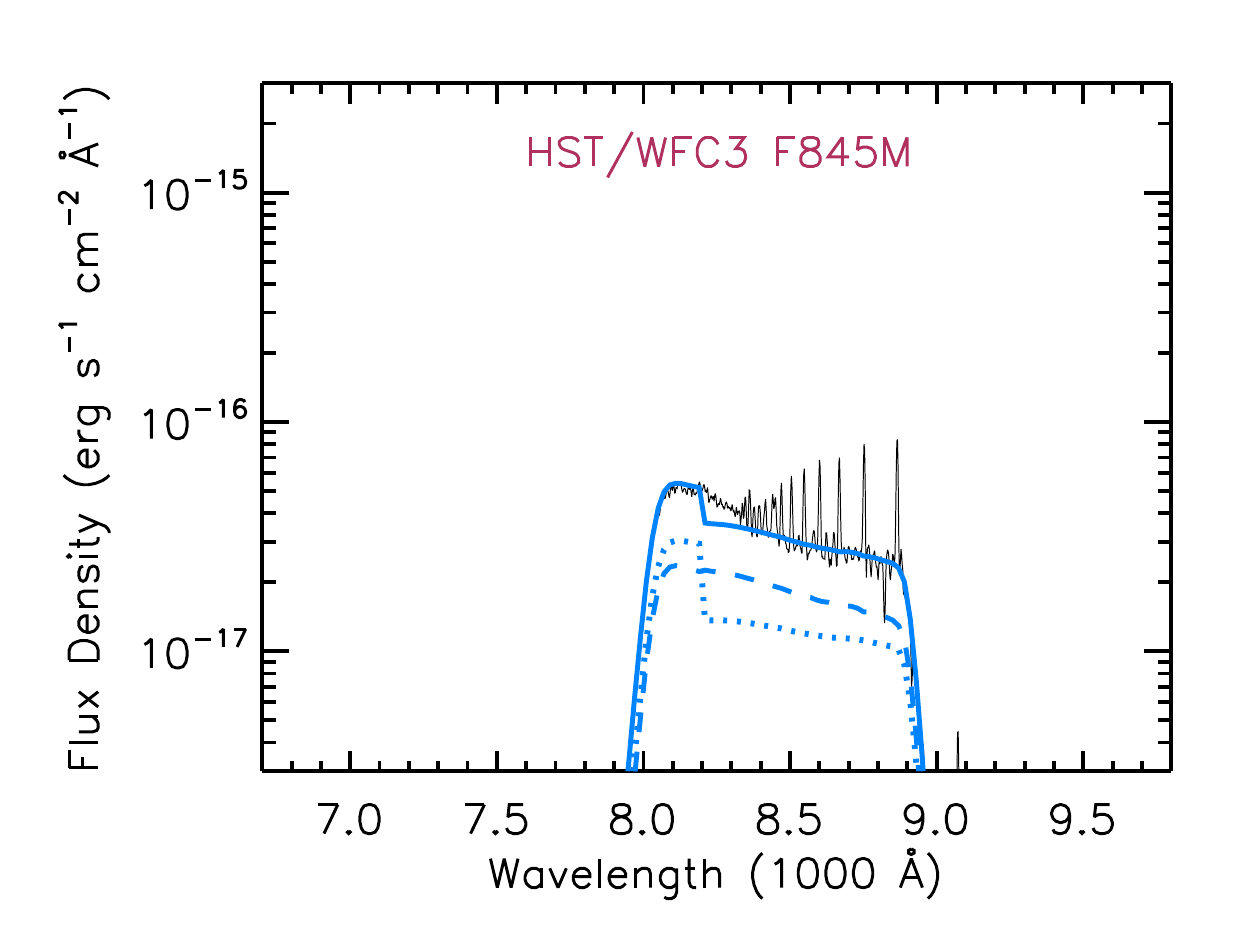}} 
\end{array}$
\end{center}
\caption{Spectrum of Source 26 multiplied by the \hst/WFC3
filters in the $V$- and $I$-band wavelength regions listed in Table
\ref{tab_contributions}.  The best-fitting Starburst99 model SED from Figure \ref{plot_seds}
is shown as a solid line.  Individual contributions from the stellar
and nebular continua are shown as dashed and dotted lines, respectively.
The impact of emission lines on the flux through various filters
is given in Table \ref{tab_contributions}.  The F555W and F814W fluxes
are most affected by line emission.
\label{plot_wfc3}}
\end{figure*}

We determine the relative contributions of stellar continuum, nebular continuum and emission lines
to the total observed flux in the F814W filter as follows.  The contributions from the stellar and
nebular continua are estimated by multiplying the models for these components with the filter throughput
curve and integrating over the bandpass.  The total flux from emission lines in this filter is determined
by multiplying the observed spectrum with the F814W filter throughput curve and measuring
the flux of each line in the resulting spectrum.
The relative contributions of each component are found by dividing by the total observed flux
in the F814W filter (obtained by convolving the observed spectrum with the throughput curve).
We find that the relative contributions of the stellar continuum, nebular continuum and emission lines
to the total observed flux of Source 26 through the F814W filter are approximately 40\%,
40\%, and 20\%, respectively.  {\it Clearly, nebular emission contributes a significant fraction ($\sim$~60\%)
of the total flux in the F814W filter and should not be neglected.}  The combination
of nebular continuum and emission lines can account for the F814W $I$-band excess found by
\citet{Reines08a}.  This result is consistent with their prediction that the origin of the excess
should not affect clusters older than $\sim$~5 Myr.

Relative contributions to the total flux of Source 26 through other \hst\ filter/instrument combinations
are listed in Table \ref{tab_contributions}.  Throughput curves were obtained using SYNPHOT,
the synthetic photometry package distributed by the Space Telescope Science Data Analysis
System (STSDAS).  The total flux and the flux of
the stellar and nebular continua are found in the same way as described above.  The line
contributions, however, are simply estimated by subtracting the continuum contributions from the
total observed flux.  Plots of the observed spectrum of Source 26 multiplied by the
\hst\ WFC3 filters listed in Table \ref{tab_contributions}
are shown in Figure \ref{plot_wfc3}.  We find that the relative contribution of the nebular
gas (continuum plus lines) to the total flux of Source 26 in any of the \hst\ $I$-band filters listed in
Table \ref{tab_contributions} is {\it at least 50\%}.  In the $V$-band, roughly 70\% of the F555W
flux comes from ionized gas.  Younger clusters will have an even larger contribution from ionized gas.
Relative contributions to the SDSS fluxes of Source 26 are also listed in Table \ref{tab_contributions}
using the throughput curves available online.\footnote{http://www.sdss.org/dr6/instruments/imager/}

We emphasize that the quantitative results listed in Table \ref{tab_contributions} are specific
to Source 26.  Relative contributions will vary from source to source, depending on (at a minimum)
age and metallicity \citep[e.g. our Figure \ref{plot_gasfrac} and][]{Anders03}.  However, it is
evident that the impact of nebular emission on broad-band fluxes can be quite large and should
not be readily dismissed.

\subsection{The Impact on Magnitudes, Colors and Derived Properties}

The ionized gas associated with a young stellar population can provide
a significant contribution to the total observed flux in a given filter
(\S\ref{sec_contributions}).  The nebular emission, therefore, can
have large impact on the physical properties derived from magnitudes
and broad-band colors.

\begin{figure*}
\begin{center}$
\begin{array}{cc}
{\includegraphics[width=3.4in]{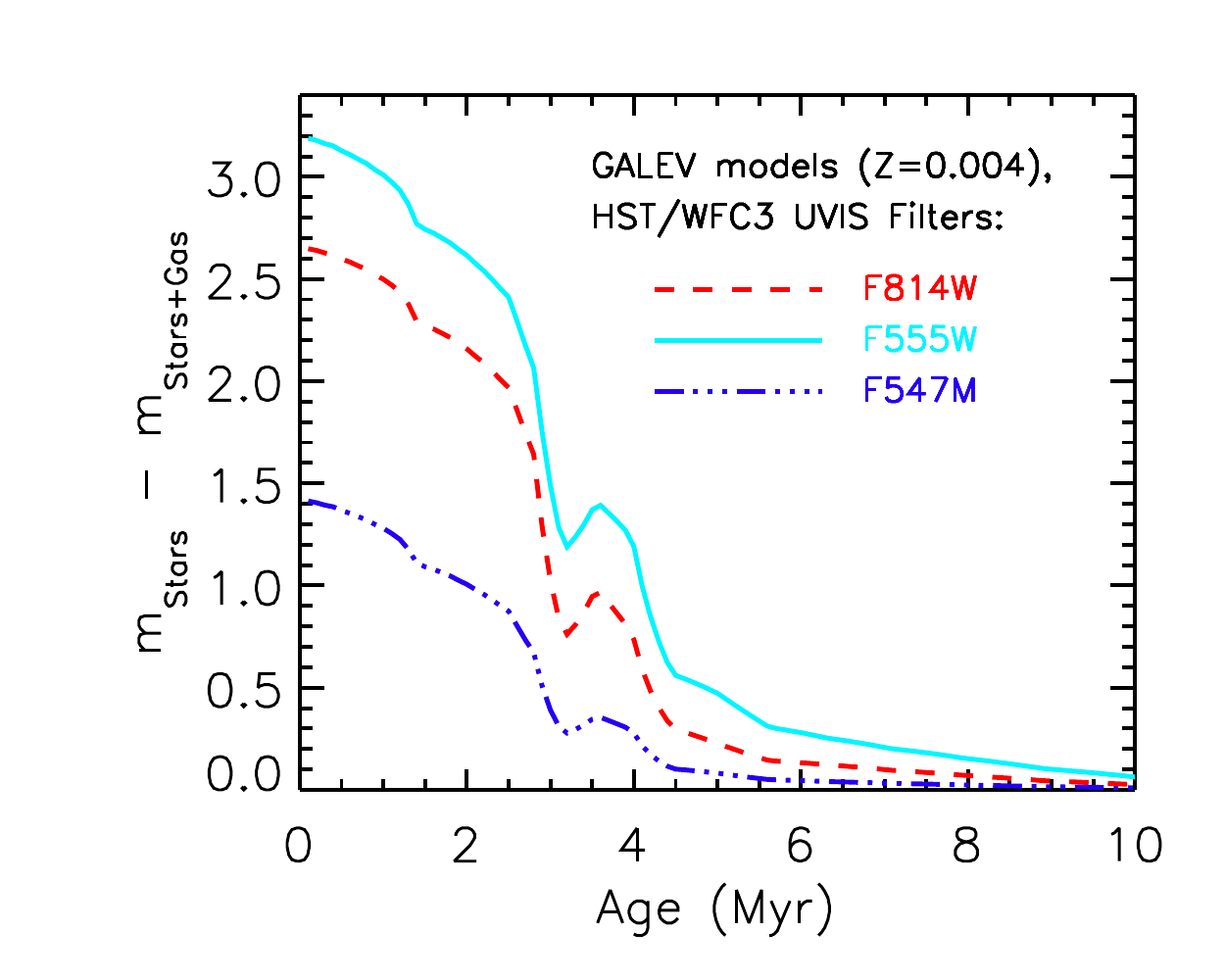}} &
{\includegraphics[width=3.4in]{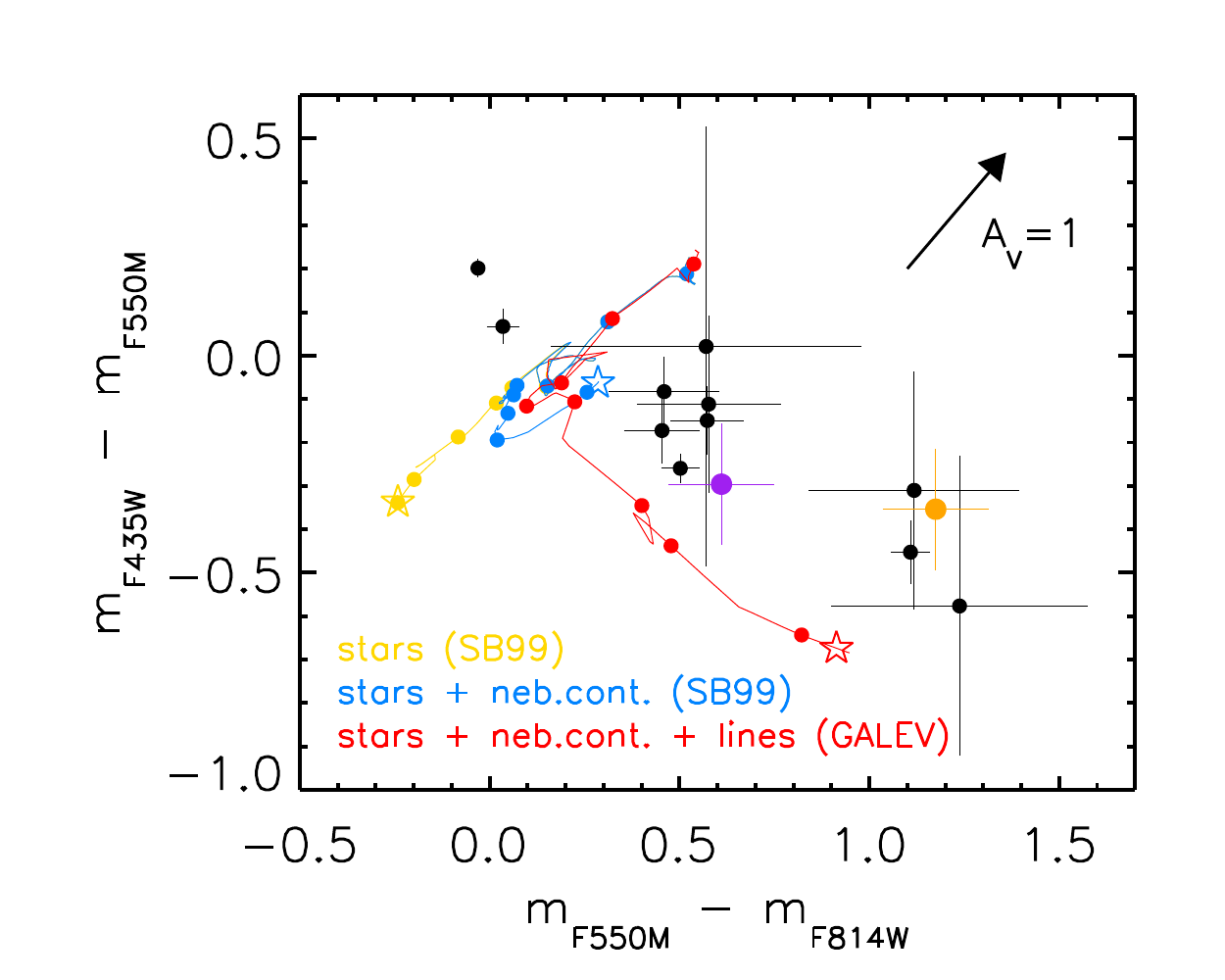}}
\end{array}$
\end{center}
\caption{{\it Left:} Magnitude difference as a function
of age between a pure stellar population and one of the same mass but including
ionized gas emission (continuum and lines).  The GALEV model SEDs described in Section
\ref{sec_galev} were convolved with the \hst/WFC3 F814W (broad $I$), F555W (wide $V$), and
F547M (narrow $V$) filters to produce this plot.
{\it Right:} Color-color diagram using the \hst/ACS WFC F435W, F550M and F814W filters
and the data of \citet{Reines08a}.  Sources 26 and 30 are shown as large purple and orange dots,
respectively.  Model evolutionary tracks are shown for a Starburst99 model with stellar continuum only
(yellow), a Starburst99 model with stellar and nebular continuum (blue), and a GALEV
model with stellar continuum, nebular continuum, and emission lines (red).
The tracks extend to an age of 50 Myr.  An age of 1 Myr is indicated by a
star and the dots along the tracks indicate ages of 2, 3, 4, 5, 7, 10, 25 and 40 Myr.
Observed and model colors are for the \hst\ filters with VEGAMAG zeropoints.
\label{plot_magcolor}}
\end{figure*}

Using the GALEV models described in Section
\ref{sec_galev}, we have computed the magnitude difference as a function
of age between a pure stellar population and one of the same mass but including
ionized gas emission (continuum plus lines) for the \hst/WFC3 F814W (broad $I$), F555W (wide $V$), and
F547M (narrow $V$) filters.  The results are shown in Figure \ref{plot_magcolor}.
For a 3 Myr old cluster, including the gaseous emission results in magnitudes that
are brighter by 1.0 (F814W), 1.5 (F555W), and 0.4 (F547M) mag.  Neglecting nebular emission
and assuming all of the flux comes from the stars would therefore lead
to overestimates of the stellar mass by factors of 2.5 (F814W), 4.0 (F555W),
and 1.4 (F547M).  From Figure \ref{plot_wfc3}, we can see that significant
contributions from both nebular continuum and lines 
are present in the F814W filter, whereas the F555W flux is most affected
by strong lines (e.g. \hbeta\ and \OIII).  The F547M flux
is affected by the nebular continuum but does not contain strong line emission.

We obtain similar results from modeling the SED of Source 26
and determining the relative contributions of
the stellar continuum, nebular continuum and emission lines to the total
flux through various filters (\S\ref{sec_contributions}).  Mass (over)estimates
of this YMC (age $\approx 3$ Myr, $Z \approx 0.004$),
assuming all of the flux in a given filter is due only to stars, can be obtained
by taking the inverse of the relative contribution of the stellar continuum to the
total flux (the third column in Table \ref{tab_contributions}).
For example, neglecting the effects of nebular emission in the ACS/WFC F555W filter
would result in a mass $3.7 \times$ larger than the mass obtained by properly
accounting for the gas contribution in this filter. 

Since the magnitudes of young stellar populations can be considerably brightened by
the presence of ionized gas, colors can also be significantly affected.  For example,
the work of \citet{Reines08a} found an apparently anomalous color-color diagram in
which the radio-detected YMCs in \ngc\ with the bluest \hst/ACS WFC [F435W$-$F550M] colors had the reddest
[F550M$-$F814W] colors, contradictory to the Starburst99 model evolutionary track
for a pure stellar continuum.  We have reproduced this color-color diagram
but have now included a Starburst99 model evolutionary track
including nebular continuum, and a GALEV track including nebular continuum plus
emission lines (Figure \ref{plot_magcolor}).  Only the GALEV model is able to reproduce the trend seen in
the data, signifying that emission lines have a large effect on these colors
at young ages ($\lesssim 5$ Myr).  Although the F550M filter is essentially line-free, the
F814W and F435W filters contain a number of emission lines.
The forbidden \SIII\ lines at 9532 and 9069 \AA\ are the most prominent in the F814W filter,
but many others are also present (see Figure \ref{plot_f814w}).  The F435W filter
contains line emission from H$\gamma$, H$\delta$, H$\epsilon +$\NIII\ $\lambda 3970$,
H$\zeta +$\HeI\ $\lambda 3889$, \NIII\ $\lambda 3869$, and \OII\ $\lambda 3727$
\citep[see our Figure \ref{plot_spectra} and][]{Anders03}.  

It is worth noting that the [F435W$-$F550M] versus [F550M$-$F814W] color-color diagram
may prove to be a useful diagnostic for identifying YMCs with ages $\lesssim 5$ Myr.
It is also interesting that the extinction vector in Figure \ref{plot_magcolor} is perpendicular
to the GALEV model track for ages $\lesssim 5$ Myr.  In other words, there is no
degeneracy in this color space between age and extinction at young ages.  Therefore, rough
estimates of both age and extinction could be obtained from this color-color plot.  However,
we emphasize that the line strengths of any given source may vary from the ones input
into the GALEV models which could potentially lead to large uncertainties in the derived properties.

As a check, we look at Sources 26 and 30 (indicated in Figure \ref{plot_magcolor}
by purple and orange dots respectively).  The ages inferred from their colors are consistent
with those obtained from their \ha\ equivalent widths and from SED
fitting; approximately 3 Myr for Source 26 and 1-2 Myr for Source 30.
The visual extinction of Source 26 also appears to be consistent with our previous estimates of
$\sim$ 0.4 magnitudes.  The extinction of Source 30 inferred from Figure \ref{plot_magcolor} ($\approx$ 1
mag), on the other hand, is half a magnitude larger than that obtained from SED fitting 
to the photometry with minimal line emission (Figure \ref{plot_seds}).\footnote{It is not clear
that optically derived \av's are accurate measures of the extinction at such young ages, since a
significant fraction of the cluster may be heavily embedded and visibly obscured.}
This difference is likely the result of different line intensities
between the models and the data.  The GALEV models, however, at least qualitatively 
explain these colors of very young star clusters, which neither of the Starburst99 models can do. 

\section{CONCLUSIONS}\label{sec_summary}

This paper examines the importance of nebular continuum and line emission in observations of young
massive stellar populations.  We have obtained spectroscopy ($\lambda \sim 3800-9800$ \AA) of
two young ($\lesssim 3$ Myr) star clusters in the nearby starburst galaxy \ngc\ and supplement
these data with archival \hst\ and SDSS photometry of the clusters.  We compare our data to
the Starburst99 and GALEV evolutionary synthesis models to estimate the physical
properties of the clusters and determine the impact of ionized gas emission (line and
continuum) on broad-band photometry.  Our main results are summarized below:

\begin{enumerate}

\item{The {\it combination} of nebular continuum and line emission
can account for the F814W $I$-band excess found by \citet{Reines08a} in
their study of radio-detected YMCs in \ngc.  This result is consistent
with their prediction that the origin of the excess should not affect
clusters older than $\sim$~5 Myr.}

\item{At very young ages ($\lesssim 3$ Myr), nebular continuum emission from ionized gas
can {\it rival} the stellar luminosity of YMCs at optical wavelengths.  
The relative contribution from the nebular continuum emission is largest in the
$U$- and $I$-bands, where the Balmer (3646 \AA) and Paschen jumps (8207 \AA) are located.}

\item{Nebular line emission is significant in many commonly used broad-band \hst\ filters
including the F814W $I$-band, the F555W $V$-band and the F435W $B$-band.  Emission lines
(in addition to nebular continuum) are required to explain the distribution of radio-detected 
YMCs in \ngc\ in the [F435W$-$F550M] versus [F550M$-$F814W] color-color diagram first presented
in \citet{Reines08a}.}

\item{SED fitting to broad-band photometry of YMCs is most successful when the photometric
bands do not contain significant line emission.  In this case, evolutionary synthesis
models only containing {\it continuum} emission from stars and ionized gas are adequate.  
If the photometric bands include significant line emission, however, models including
emission lines perform much better than those that do not.}

\item{The impact of ionized gas emission on broad-band fluxes, magnitudes and colors can be
large and should not be readily dismissed.  We urge caution when comparing observations
of young ($\lesssim 5$ Myr) clusters to synthesis models since nebular continuum and line
emission can significantly affect inferred properties such as ages, masses and extinctions.}

\end{enumerate}

\acknowledgments

We thank Ralf Kotulla for providing us with the custom-run GALEV models
used in this work.  We also thank the anonymous referee for numerous
useful comments and suggestions which led to the overall improvement
of this paper.  A.E.R. appreciates helpful discussions with Adolf Witt,
Yuri Izotov and Crystal Brogan.  A.E.R. also gratefully acknowledges support from 
a NASA Earth and Space Science Fellowship, the Virginia Space Grant
Consortium and the University of Virginia through a Governor's Fellowship.
K.E.J. gratefully acknowledges support provided in part by NSF
through CAREER award 0548103 and the David and Lucile Packard
Foundation.  Support for program \#AR09934 was provided in part by NASA
through a grant from the Space Telescope Science Institute, which is operated
by the Association of Universities for Research in Astronomy, Inc., under NASA
contract NAS 5-26555.

\end{document}